\documentclass[aps,prb,twocolumn,floats]{revtex4}
\usepackage{color,ulem,verbatim,bbold,float,wrapfig}
\usepackage{amssymb,amsbsy,amsmath,mathrsfs}
\usepackage{epsfig,subfigure}
\usepackage{graphicx}
\usepackage{times}
\usepackage{color}
\usepackage{subfigure}
\usepackage{setspace}
\usepackage{bm}
\newcommand\bea{\begin{eqnarray}}
\newcommand\eea{\end{eqnarray}}
\newcommand\beq{\begin{equation}}
\newcommand\eeq{\end{equation}}
\newcommand\bib{\bibitem}

\newcommand{\noi}{\noindent}
\newcommand{\non}{\nonumber}
\newcommand{\al}{\alpha}
\newcommand{\de}{\delta}

\newcommand{\lm}{\lambda}
\newcommand{\si}{\sigma}
\newcommand{\ta}{\theta}

\newcommand{\pa}{\partial}
\newcommand{\la}{\langle}
\newcommand{\ra}{\rangle}

\newcommand{\bra}[1]{\langle #1|}
\newcommand{\ket}[1]{|#1\rangle}

\usepackage{amsmath}
\usepackage{amssymb}
\usepackage{float}
\usepackage{graphicx}
\usepackage{caption}
\usepackage{subfigure}
\usepackage{dsfont}

\begin{document}

\title{Transport in a thin topological insulator with potential and magnetic
barriers}

\author{Adithi Udupa$^1$, K. Sengupta$^2$ and Diptiman Sen$^1$}

\affiliation{$^1$Center for High Energy Physics, Indian Institute of Science,
Bengaluru 560012, India \\
$^2$Theoretical Physics Department, Indian Association for the Cultivation
of Science, Jadavpur, Kolkata 700032, India}

\begin{abstract}

We study transport across either a potential or a magnetic barrier
which is placed on the top surface of a three-dimensional
thin topological insulator (TI). For such thin TIs, the top
and bottom surfaces interact via a coupling $\lm$ which
influences the transport properties of junctions constructed out of
them. We find that for junctions hosting a potential barrier,
the differential conductance oscillates with the barrier
strength. The period of these oscillations doubles as the coupling
$\lm$ changes from small values to a value close to the energy of
the incident electrons. In contrast, for junctions with a magnetic
barrier, the conductance approaches a nonzero constant as the
barrier strength is increased. This feature is in contrast to the
case of transport across a single TI surface where the conductance
approaches zero as the strength of a magnetic barrier is increased.
We also study the spin currents for these two kinds of barriers; in
both cases, the spin current is found to have opposite signs on the
top and bottom surfaces. Thus this system can be used to split
applied charge currents to spin currents with opposite spin
orientations which can be collected by applying opposite
spin-polarized leads to the two surfaces. We show that several of
these features of transport across finite width barriers can be
understood analytically by studying the $\de$-function barrier limit.
We discuss experiments which may test our theory.
\end{abstract}

\maketitle

\section{Introduction}
\label{sec1}

Three-dimensional topological insulators have been extensively
studied for the last several years both
theoretically~\cite{kane1,bern,kane2,moore,hasan1,qi2} and
experimentally~\cite{chen,cheng,hsieh1,hsieh2,xia,roushan}. A
topological insulator (TI) is a material which is gapped in the bulk
and has gapless states at all the surfaces which have a Dirac-type
linear energy-momentum dispersion and are protected by time-reversal
symmetry. Examples of such materials include Bi$_2$Se$_3$ and
Bi$_2$Te$_3$. The bulk topological aspects of these
TIs can be characterized by four integers $\nu_0$ and $\nu_{1,2,3}$
\cite{kane2}. The first integer $\nu_0$ classifies these TIs as
strong ($\nu_0=1$) or weak ($\nu_0=0$), while the others,
$\nu_{1,2,3}$, characterize the time-reversal invariant momenta at
which the bulk Kramer pair bands cross: $\vec L_0 = (\nu_1 \vec b_1,
\nu_2 \vec b_2, \nu_3 \vec b_3)/2$, where $\vec b_{1,2,3}$ are the
reciprocal lattice vectors. The strong topological insulators are
robust against the presence of time-reversal invariant perturbations such
as nonmagnetic disorder or lattice imperfections. It is well-known that
\cite{kane2,qi2,hasan1} the surface of a strong TI has an odd number
of Dirac cones. The positions of these cones are determined by the
projection of $\vec L_0$ on to the surface Brillouin zone. The
number of these cones depends on the nature of the surface; for
example, for materials such as ${\rm HgTe}$ and ${\rm Bi_2 Se_3}$,
surfaces with a single Dirac cone at the center of the two-dimensional
Brillouin zone have been found \cite{hasan1,hsieh1,hsieh2}.

The effective Dirac Hamiltonians governing the surface states can be
derived starting from the bulk continuum
Hamiltonian~\cite{liu2,zhang3,deb1}. The surface states are known to
exhibit spin-momentum locking in which the directions of spin
angular momentum and linear momentum lie in the same plane and are
perpendicular to each other~\cite{hsieh4}. Several interesting
properties of these surface Dirac electrons have been studied. These
include proximity effects between an s-wave superconductor and the
surface states and the consequent appearance of Majorana
states~\cite{fu2}. Various properties of junctions between different
surfaces of TIs have been studied in
Refs.~\onlinecite{deb2,wickles,biswas,alos,sitte,apalkov,habe}.
Junctions of surfaces of a TI with normal metals, magnetic materials
and superconductors have also been studied~\cite{modak,soori,nuss}.
The effects of potential, magnetic and superconducting barriers on
the surface of a TI have been studied in Refs.~\onlinecite{mondal}
and \onlinecite{peeters}. Spin-charge coupled transport on the surface of 
a TI has been studied in Ref.~\onlinecite{burkov}, leading to interesting
magnetoresistance effects. Magnetic textures, such as domain walls and 
vortices, in a ferromagnetic thin film deposited on the surface of a TI have 
been examined in Ref.~\onlinecite{nomura}. The dynamics of magnetization 
coupled to the surface Dirac fermions has been investigated theoretically 
in Ref.~\onlinecite{yoko}. There have also been studies of transport in 
TI $p-n$ junctions in the presence of a magnetic field~\cite{ilan,dai}, 
magnetotransport in patterned TI nanostructures~\cite{moors}, and the 
effects of disorder on transport~\cite{zhou}.

More recently, several theoretical studies have been
carried out for thin films of a TI where the hybridization of the
states on the opposite surfaces of the system~\cite{shan} gives rise
to interesting phenomena. These phenomena include quantum phase transitions
in the presence of a parallel magnetic field~\cite{zyu} and the appearance
of a number of topological and nontopological phases~\cite{asmar}. It has 
been shown that a Coulomb interaction between the opposite surfaces can give 
rise to a topological exciton condensate~\cite{serad1}, and a Zeeman field 
and a proximate superconductor can then give rise to Majorana edge 
modes~\cite{serad2}. A number of other effects of finite width have been 
studied in Refs.~\onlinecite{shen,linder,egger,kundu,shenoy,pertsova}. 
However, the transport properties of such thin TIs in the presence of
potential or magnetic barriers have not been studied before.
Motivated by the above studies, we will consider in this paper a simple
model of a TI with a coupling, characterized by a strength
$\lm$, between the top and the bottom surfaces; we will study
the various features of electronic transport in such a system when a
potential or magnetic barrier is applied on one of the surfaces.

The main results that arise out of our study can be
summarized as follows. First, we show that for junctions with a
potential barrier on the top surface, the tunneling conductance $G$
of the junctions oscillates with the barrier strength. The period of
these oscillations can be tuned by changing $\lm$; it doubles as
$\lm$ is increased from zero to a value close to the incident
energy of the Dirac electrons on the surface. Second, for a magnetic
barrier, we find that the tunneling conductance reaches a nonzero
and $\lm$-dependent value as the barrier strength is increased.
This is in sharp contrast to the behavior of $G$ for a single TI surface 
where it approaches zero with increasing magnetic barrier strength. 
Third, for both potential and magnetic barriers, we compute the spin 
current for the top and the bottom surfaces and demonstrate
that they always have opposite signs which implies opposite spin 
polarizations. The origin of this can be traced to the opposite helicities 
of the Dirac electrons on these two surfaces. Our results thus indicate
that these junctions may be used to split an applied charge current
into two spin currents with opposite directions of spins. These spin
currents may be collected, for example, by connecting spin-polarized
leads to the top and the bottom surfaces.

The plan of this paper is as follows. In Sec.~\ref{sec2}, we discuss
a model of the top and bottom surfaces of a TI such as Bi$_2$Se$_3$,
with a coupling $\lm$ between the two surfaces. We will then present
the form of the Hamiltonian when a potential or magnetic barrier of
finite width is applied on the top surface. Next, in
Sec.~\ref{sec3}, we will discuss the forms of the wave functions in
the two regions where there are no barriers and the matching
conditions at the interfaces between these regions and the middle
region where there is a barrier. We will introduce a basis in which
the transmitted charge currents can be calculated most easily, and
we will present expressions for the transmitted charge and spin
currents. This will be followed by Sec.~\ref{sec4} where we will
discuss the case of $\de$-function barriers. Such barriers induce
discontinuities in the wave functions. This problem turns out to be
easier to study than the case of finite width barriers since the
matching conditions involve four equations instead of eight
equations. We obtain analytical expressions for the reflection and
transmission amplitudes in some special cases. Next, in Sec.~\ref{sec5},
we will study the case of a potential barrier with a finite width and present
numerical results as a function of various parameters such as $\lm$, the
angle of incidence $\ta$, and the barrier strength $V_0$. We
point out certain symmetries of the transmission probabilities under
$\ta \to \pi - \ta$. We also study the transmitted charge and spin
currents at the top and bottom surfaces separately. This is followed
by Sec.~\ref{sec6}, where we will present numerical results for the
case of a magnetic barrier with a finite width. Finally, in
Sec.~\ref{sec7}, we will summarize our main results, suggest possible
experiments which can test our theory, and conclude.

\section{Model of top and bottom surfaces}
\label{sec2}

As mentioned above, a three-dimensional TI has gapless surface states on
all its surfaces and
the eigenstates of the Hamiltonian at the top and bottom surfaces
exhibit spin-momentum locking. Namely, on a given surface, the directions of
the linear and spin angular momentum are perpendicular to each other,
and the relation between the two is opposite on the top and bottom
surfaces. To show this, we begin with the bulk Hamiltonian of the system
near the $\Gamma$ point. This is known to have the form~\cite{qi2}
\begin{equation} H_{\vec{k}} ~=~ m\tau^z ~+~ \hbar v_z\tau^y k_z ~+~ \hbar v
\tau^x (\si^x k_y -\si^y k_x). \label{ham0} \end{equation}
In Bi$_2$Se$_3$, which is a well-known TI, the parameters in Eq.~\eqref{ham0}
have the values $m = 0.28$ eV, $\hbar v_z = 0.226$ eV-nm, and 
$\hbar v = 0.333$ eV-nm. (We will henceforth set $\hbar = 1$ unless 
explicitly mentioned). The energy-momentum dispersion is found by solving the 
equation $H \psi = E \psi$, where $\psi$ is a four-component wave function 
given by
\beq \psi ~=~ e^{i(k_x x+k_y y+k_z z - E t)} ~\begin{pmatrix}
\phi_1 \\
\phi_2 \\
\phi_3 \\
\phi_4\\
\end{pmatrix}, \eeq
where two of the components represent wave functions of electrons
localized on different orbitals (for example, Bi and Se in the
material Bi$_2$Se$_3$) and the other two components represent the spin
degrees of freedom (up and down). The matrices $\tau^a$ act on the
pseudospin components and the matrices $\si^a$ act on the spin components.
[We will work in a basis in which $\tau^z$ and $\si^z$ are diagonal matrices;
the diagonal entries of the two matrices are given by $\tau^z = (1,1,-1,-1)$
and $\si^z = (1, -1, 1, -1)$].
Since the four matrices appearing in Eq.~\eqref{ham0}, $\tau^z, ~\tau^y, ~
\tau^x \si^x$ and $\tau^x \si^y$ anticommute with each other, the Hamiltonian
has the form of an anisotropic Dirac equation in three dimensions; the
dispersion is given by
\beq E ~=~ \pm ~\sqrt{m^2 ~+~ v_z^2 k_z^2 ~+~ v^2 (k_x^2 ~+~ k_y^2)}.
\label{disp} \eeq
At the $\Gamma$ point, there is a gap equal to $2m$ between the positive
and negative energy bands.

To derive the Hamiltonian on the surface from the bulk
Hamiltonian~\cite{zhang3,deb1}, we consider the top surface to be at
$z=0$ with the region with $z < 0$ being the TI and
$z > 0$ being the vacuum. Further, we will assume $m$ in Eq.~\eqref{ham0}
to be a function of $z$; in the vacuum, we take $m$ to be large and negative,
while in the interior of the TI (with $z < 0$), $m$ is a
positive constant ($0.28$ eV in Bi$_2$Se$_3$). Since the momentum along $z$
is not a good quantum number, we replace $k_z \to -i \pa / \pa z$.
Writing the bulk Hamiltonian as a sum, $H_{\vec{k}}= H_0+H_s$ with
\bea H_0 &=& m\tau^z - i v_z \tau^y \dfrac{\pa}{\pa z}, \non \\
H_s &=& v\tau^x (\si^x k_y - \si^y k_x), \eea
acting on the wave function $\psi(x,y,z)= e^{ik_x x+ik_y y} ~f(z) ~\phi,$
where $\phi$ is a four-component column. (For convenience, we will not write 
the time-dependent factor $e^{-iEt}$ any longer).
For ${\vec k} = 0$, we know that $H_0$ has a zero energy eigenstate
localized near the surface, namely, $H_0 \psi= 0$, where $f(z)$ has the form
\beq f(z) ~=~ e^{\frac{1}{v_z}\int_0^z dz' m(z')}. \label{fz} \eeq
This gives the condition $(\tau^z-i\tau^y)\psi=0$. This implies that
$(\tau^z+i\tau^y) (\tau^z-i\tau^y)\psi=0$ giving $\tau^x\psi = \psi$. Since 
$H_{s}$ commutes with $\tau^x$ and $H_{s}\psi= E\psi$, we find from the above
that $v(k_y\si^x - k_x\si^y)\psi= E\psi$ with $E= \pm
\sqrt{v^2(k_x^2+k_y^2)}$. Thus the Hamiltonian on the top surface is
\begin{equation} H_{top}= v(\si^x k_y - \si^y k_x). \label{hamt} \end{equation}
Similarly on the bottom surface, we get
\begin{equation} H_{bottom}= - v(\si^x k_y - \si^y k_x), \label{hamb}
\end{equation}
again with $E= \pm v \sqrt{k_x^2+k_y^2}$. We note that the Hamiltonians in
Eqs.~\eqref{hamt} and \eqref{hamb}) have opposite signs. This leads to 
opposite forms
of spin-momentum locking on the two surfaces; an electron with positive
energy and moving in the $\hat k$ direction on the top (bottom) surface has
a spin pointing in the $-{\hat z} \times {\hat k}$ (${\hat z} \times {\hat k}$)
direction, respectively.

If the separation between the two surfaces is not much larger than the decay
length of the surface states (Eq.~\eqref{fz} implies that this length is
about $v_z/m$), there will be some hybridization between the two surfaces
states. We can parametrize this by a tunneling coupling $\lm$ which has
dimensions of energy. The total Hamiltonian for the two surfaces then
becomes~\cite{shan}
\begin{equation} H_{0} = \begin{pmatrix}
H_{top} & \lm I_2 \\
\lm I_2 & H_{bottom}
\end{pmatrix}, \label{ham1} \end{equation}
where $I_2$ denotes the two-dimensional identity matrix. The value of $\lm$ can
be estimated as follows. If $w$ is the width of the material in the $\hat z$ 
direction, so that the top and bottom surfaces lie at $z=0$ and $z=-w$,
respectively, the tunneling $\lm$ between the two surfaces can be shown to be
proportional to $m e^{- mw/v_z}$. Note that for such a finite width sample,
the momentum $k_z$ of the bulk states will be quantized in units of $\pi/w$.
However, Eq.~\eqref{disp} shows that the bulk
states will continue to have a gap equal to $2m$.
Hence, they will not affect our results since we are only interested in the
contributions of the surface states which lie within the bulk gap.

\begin{figure}[H]
\centering
\hspace*{-.4cm} \includegraphics[scale=0.58]{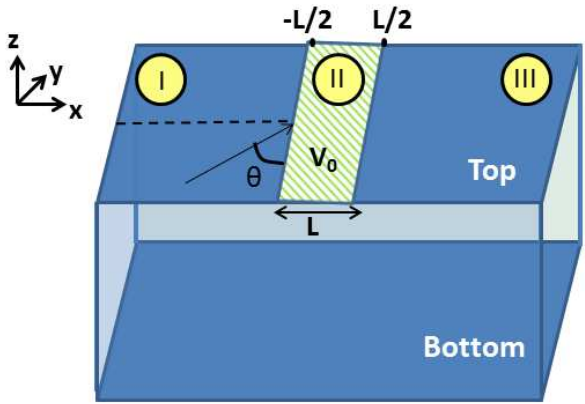}
\caption{Schematic picture of the system showing the top and bottom surfaces
of a TI, a potential barrier with strength $V_0$ and width
$L$ on the top surface (region $II$), and a wave coming in from region $I$
with an angle of incidence $\ta$.} \label{fig01} \end{figure}

We will study the effects of two kinds of barriers on the top surface. In
Sec.~\ref{sec5}, we will study what happens if the top surface has a potential
barrier which is independent of the $y$ coordinate and has the form
$V(x)= V_{0}$ in a region of width $L$. (A schematic picture of this is
shown in Fig.~\ref{fig01}). The Hamiltonian of this system is given by
\begin{equation} H_{0} = \begin{pmatrix}
H_{top}+V_{0} I_2 & \lm I_2 \\
\lm I_2 & H_{bottom}
\end{pmatrix} \label{ham2} \end{equation}
for $-L/2 < x < L/2$, and by Eq.~\eqref{ham1} for $x < -L/2$ and $x > L/2$.
In Sec.~\ref{sec6}, we will study what happens if the top surface has
a magnetic barrier of strength $V_0$ in a region of width $L$. As explained
below, we will choose the direction of the magnetization in the barrier in
such a way that the Hamiltonian in the region $-L/2 < x < L/2$ has the form
\begin{equation} H_{0} = \begin{pmatrix}
H_{top}+V_{0} \si^x & \lm I_2 \\
\lm I_2 & H_{bottom}
\end{pmatrix}. \label{ham3} \end{equation}
(In Fig.~\ref{fig01}, this corresponds to having a barrier with strength
$V_0 \si^x$ in region $II$). In both cases, our aim will be to study the 
transmitted charge and spin currents and their dependences on the various 
parameters of the system, namely, the energy $E$, the coupling between the 
two surfaces $\lm$, and the width and height of the potential barrier $L$ 
and $V_0$.

In this paper, we are assuming that the TI is in the form of a thin film 
whose top and bottom
surfaces cover a large area in the $x-y$ plane and whose thickness in 
the $z$-direction is small. In this situation, which is common for experimental 
measurements of transport, the contributions of the side surfaces are much 
smaller than those of the top and bottom surfaces and can therefore be ignored.

\section{Barrier-free regions}
\label{sec3}

In the barrier-free regions denoted as $I$ and $III$, the Hamiltonian is
\begin{equation} H_{0} = \begin{pmatrix}
0 & vke^{i\ta} & \lm & 0 \\
vke^{-i\ta} & 0 & 0& \lm \\
\lm & 0 & 0 & -vke^{i\ta}\\
0& \lm & -vke^{-i\ta}& 0 \end{pmatrix}, \label{ham13} \end{equation}
where $ke^{i\ta} = k_y + i k_x$. Thus $k= \sqrt{k_x^2+ k_y^2}$ and
$\ta = \tan^{-1} (k_x/k_y)$. Defining $E = \sqrt{v^2k^2+\lm^2}$, the
eigenvalues of the Hamiltonian in Eq.~\eqref{ham13} are \bea e_1 &=&
e_2 ~=~ -~ e_3 ~=~ - ~ e_4 ~=~ E,\eea
with corresponding eigenstates
\bea \ket{e_{1}} &=& \frac{1}{2E} ~\begin{pmatrix}
\sqrt{v^{2}k^{2}+\lm^{2}} \\ \lm+ vke^{-i\ta}\\ \lm -vke^{i\ta}\\
-\sqrt{v^{2}k^{2}+\lm^{2}} \end{pmatrix}, \non \\
\ket{e_{2}} &=& \frac{1}{2E} ~\begin{pmatrix}
\sqrt{v^{2}k^{2}+\lm^{2}} \\ -\lm+ vke^{-i\ta}\\
\lm +vke^{i\ta}\\ \sqrt{v^{2}k^{2}+\lm^{2}} \end{pmatrix}, \non \\
\ket{e_{3}} &=& \frac{1}{2E} ~\begin{pmatrix} -\sqrt{v^{2}k^{2}+\lm^{2}} \\
\lm- vke^{-i\ta}\\ \lm -vke^{i\ta}\\
\sqrt{v^{2}k^{2}+\lm^{2}} \end{pmatrix}, \non \\
\ket{e_{4}} &=& \frac{1}{2E} ~\begin{pmatrix}
\sqrt{v^{2}k^{2}+\lm^{2}} \\ \lm+ vke^{-i\ta}\\ -\lm -vke^{i\ta}\\
\sqrt{v^{2}k^{2}+\lm^{2}} \end{pmatrix}. \eea

\subsection{Wave functions and boundary conditions}
\label{sec3a}

In the presence of a potential or a magnetic barrier on the top surface,
the reflection and transmission amplitudes can be calculated as follows.

On the top surface, we have three regions: the incident region, the potential
region of width $L$, and the transmitted region. Since the Hamiltonian has
the Dirac form (i.e., first order in the spatial derivatives), we must match
the wave functions (but not their derivatives) at the boundaries between the
incident region $I$ and the barrier region (labeled as $II$), and between the
barrier region $II$ and the transmitted region $III$. Let these boundaries be
at $x=-L/2$ and $x=L/2$. We then have the following wave functions in the
three regions.

In the incident region $I$, we consider an incident wave with 
positive energy, $E=\sqrt{v^{2}k^{2}+\lm^{2}}$, and one of the 
eigenstates, say, $\ket{e_{3}}$. There will then be two possible reflected 
wave functions with the same energy $E$ and amplitudes $r_{1}$ and $r_{2}$. 
The incident and reflected waves are given by
\bea \ket{\psi_{in}} &=& \ket{e_{3}}e^{i(k_x x+k_y y)}, \non \\
\ket{\psi_{ref}} &=& (r_{1}\ket{e_{-3}} ~+~ r_{2}\ket{e_{-4}})~
e^{i(-k_x x+k_y y)}, \eea
where $\ket{e_{3}}$ and $\ket{e_{4}}$ have been defined earlier, and
$\ket{e_{-3}}$ and $\ket{e_{-4}}$ can be obtained from those by changing
$k_x \to -k_x$ since these are reflected wave functions. The total wave 
function in this region is $\ket{\psi_I} = \ket{\psi_{in}} + \ket{\psi_{ref}}$.

In the transmitted region $III$, we have two possible wave functions, with 
amplitudes $t_{1}$ and $t_{2}$. Thus
\bea \ket{\psi_{III}} ~=~ (t_{1}\ket{e_{3}} ~+~ t_{2}\ket{e_{4}})~
e^{i(k_x x+k_y y)}. \eea

We now turn to the barrier region $II$.
Since the barrier is independent of the $y$ coordinate, the momentum
in the $\hat y$ direction, $k_y$, and, of course, the energy $E$ will
be the same in all the regions. However, the momentum in the $\hat x$ direction
will generally be different in region $II$ as compared to regions $I$ and
$III$. In region $II$, therefore, we will have four different eigenstates
having amplitudes $C_{1}, C_{2}, C_{3}$ and $C_{4}$. Namely, we have
\bea \ket{\psi_{II}} &=& C_{1}\ket{e_{1}'}e^{i(k_{x1}'x+k_y y)} ~+~ C_{2}
\ket{e_{2}'}e^{i(k_{x2}'x+k_y y)} \non \\
&& +~ C_{3}\ket{e_{3}'} e^{i(k_{x3}'x +k_y y)} ~+~ C_{4}\ket{e_{4}'}
e^{i(k_{x4}'x+k_y y)}, \non \\
&& \eea
where $k_{xi}'$ denotes the four possible values of the momentum in region
$II$; these four values and the corresponding wave functions $\ket{e_{i}'}$
depend on the nature of the barrier (potential and magnetic), and we will
present them in Secs.~\ref{sec5} and \ref{sec6}.

Applying the matching conditions at the boundaries, we obtain
\bea \ket{\psi_{I}} &=& \ket{\psi_{II}} ~~~~\text{at}~~~ x=-L/2, \non \\
\ket{\psi_{II}} &=& \ket{\psi_{III}} ~~~\text{at}~~~ x=L/2. \eea
There are thus eight unknowns, $r_{1},r_{2},t_{1},t_{2},C_{1},C_{2},C_{3},C_4$,
and we have eight equations from matching the four-component wave
functions at $x= \pm L/2$. We can therefore solve for the unknowns by writing
the eight-dimensional columns $A= (r_{1},r_{2},t_{1},t_{2},C_{1},C_{2},C_{3},
C_{4})^{T}$ and $B=(\psi_{in},0,0,0,0)^{T}$ which are related by a matrix $M$
such that $MA=B$; the elements of $M$ are obtained by writing the amplitudes
from the various equations above. We can then find the unknowns numerically.

\subsection{Basis of eigenstates of $\tau^x \si^z$}
\label{sec3b}

Once we obtain the transmission amplitudes $t_{1}$ and $t_{2}$ after solving
for the column $A$, the transmitted current and its properties can be studied.
This calculation becomes simpler if we make a change of basis as follows.
We observe that the Hamiltonian in the barrier-free regions $I$ and $III$
can be written as
\begin{equation} H_0 ~=~ v \tau^z (\si^x k_y - \si^y k_x) ~+~
\lm \tau^x. \end{equation}
We note that $\tau^x \si^z$ commutes with $H_0$. Next, we see that
\beq \tau^x \si^z ~=~ \begin{pmatrix}
0 & 0 & 1 & 0 \\
0 & 0 & 0 & -1 \\
1 & 0 & 0 & 0 \\
0 & -1 & 0 & 0 \end{pmatrix} \eeq
has eigenvalues $\pm 1$ (both doubly degenerate) and corresponding
eigenstates of the form
\beq \ket{1} ~=~ \begin{pmatrix}
a \\ b \\ a \\ -b
\end{pmatrix} ~~~{\rm and}~~~
\ket{-1} ~=~ \begin{pmatrix}
a' \\ b' \\ -a' \\ b' \end{pmatrix}. \label{form} \eeq
Since we can find simultaneous eigenstates of $\tau^x \si^z$ and $H_0$ in
regions $I$ and $III$, we look for eigenstates of $H_{0}$
which have the forms given in Eq.~\eqref{form} and which satisfy
\bea H_{0} \ket{1} &=& E\ket{1}, \non \\
H_{0} \ket{-1} &=& E\ket{-1}, \eea
with energy $E = \sqrt{v^2 k^2 + \lm^2}$. We find that the eigenstates
for the incident waves have the form
\bea \ket{1_{in}} &=& \dfrac{1}{2 \sqrt{E}} \begin{pmatrix}
\sqrt{E+\lm} \\ \sqrt{E-\lm}e^{-i\ta} \\ \sqrt{E+\lm} \\ -\sqrt{E-\lm}e^{-i\ta}
\end{pmatrix}, \non \\
\ket{-1_{in}} &=& \dfrac{1}{2 ~\sqrt{E}} \begin{pmatrix}
\sqrt{E-\lm} \\ \sqrt{E+\lm}e^{-i\ta} \\ -\sqrt{E-\lm} \\ \sqrt{E+\lm}e^{-i\ta}
\end{pmatrix}. \label{sim1} \eea
We can see that in the $\lm \to 0$ limit, there are two linear combinations of 
the above wave functions which have components only at the top and bottom 
surfaces, respectively.

To obtain the reflected waves, we change $k_x \to -k_x$, i.e., $\ta \to - 
\ta$. This gives
\bea \ket{1_{ref}} &=& \dfrac{1}{2 \sqrt{E}} \begin{pmatrix}
\sqrt{E+\lm} \\ \sqrt{E-\lm}e^{i\ta} \\ \sqrt{E+\lm} \\ -\sqrt{E-\lm}e^{i\ta}
\end{pmatrix}, \non \\
\ket{-1_{ref}} &=& \dfrac{1}{2 \sqrt{E}} \begin{pmatrix}
\sqrt{E-\lm} \\ \sqrt{E+\lm}e^{i\ta} \\ -\sqrt{E-\lm} \\ \sqrt{E+\lm}e^{i\ta}
\end{pmatrix}. \label{sim2} \eea

Choosing $\ket{1_{in}}$ to be the incident wave, we have
\bea \ket{\psi_{I}} &=& \ket{1_{in}}e^{i(k_x x+k_y)} \non \\
&& +~ (r_{1}\ket{1_{ref}} ~+~ r_{2}\ket{-1_{ref}}) ~e^{i(-k_x x+ k_y y)} \eea
in region $I$, and
\beq \ket{\psi_{III}} ~=~ (t_{1} \ket{1_{in}} ~+~ t_{2} \ket{-1_{in}})~
e^{i(k_x x+k_y y)} \eeq
in region $III$.

The advantage of working in the basis of eigenstates of $\tau^x \si^z$ is that
the charge current, when calculated in regions $I$ and $III$, will not
have any cross-terms involving $r_1, ~r_2$ and $t_1, ~t_2$. To see this,
we note that for the Hamiltonian $H_0$, the current operators can
be found using the equation of continuity and are given by
\bea J_x &=& -v \tau^z \si^{y} ~=~ \begin{pmatrix}
0 & iv & 0 & 0 \\
-iv & 0 & 0 & 0 \\
0 & 0 & 0 & -iv \\
0 & 0 & iv & 0 \end{pmatrix}, \non \\
J_y &=& v\tau^z \si^x ~=~ \begin{pmatrix}
0 & v & 0 & 0 \\
v & 0 & 0 & 0 \\
0 & 0 & 0 & -v \\
0 & 0 & -v & 0 \end{pmatrix}. \label{jxy} \eea
We see that both $J_x$ and $J_y$ commute with the operator $\tau^x \si^z$.
We will only study $J_{x}$ below. We now note that
\bea \bra{1}J_{x}\ket{-1} &=& \bra{1}J_{x}(\tau^{x}\si^z)^2 \ket{-1} \non \\
&=& \bra{1}\tau^{x}\si^z J_{x}\tau^{x}\si^z\ket{-1} \non \\
&=& - ~\bra{1} J_{x}\ket{-1}. \eea
This implies that $\bra{1} J_{x}\ket{-1} =0$; hence there will be no 
cross-terms when we calculate the expectation value of $J_{x}$ in
regions $I$ and $III$.

\subsection{Conservation of charge current in the $\hat x$ direction}
\label{sec3c}

Given the wave functions and the form of $J_{x}$, we can calculate
$\la J_{x}\ra$ in regions $I$ and $III$ and check for conservation of the
charge current. In region $I$, we have
\bea J_{x} \ket{\psi_I} &=& J_{x} [\ket{1_{in}} e^{i(k_{x}x+k_{y})} \non \\
&& + (r_{1}\ket{1_{ref}} + r_{2}\ket{-1_{ref}}) e^{i(-k_{x}x+ k_{y}y)}]. \eea
Since
\bea \bra{\psi_{I}} &=& \bra{1_{in}} e^{-i(k_{x}x+k_{y})} \non \\
&& + (r_{1}^{*} \bra{1_{ref}} + r_{2}^{*}\bra{-1_{ref}})
e^{-i(-k_{x}x+ k_{y}y)}, \eea
we find that
\beq \bra{\psi_{I}} J_{x} \ket{\psi_{I}} ~=~\dfrac{v\sqrt{E^{2}-
\lm^{2}}}{E} ~\sin \ta ~(1-|r_{1}|^{2}-|r_{2}|^{2}). \eeq
Using the relation $E^2 = v^{2}k^{2} + \lm^{2}$, we can simplify this to obtain
\beq \la J_{x} \ra_{I} ~=~ \dfrac{v^{2}k}{E} ~\sin \ta
~(1-|r_{1}|^{2}-|r_{2}|^{2}). \label{jI} \eeq
In region $III$, we calculate $\bra{\psi_{III}} J_{x} \ket{\psi_{III}}$,
and find that
\beq \la J_{x}\ra_{III} ~=~ \dfrac{v^{2}k}{E} ~\sin
\ta ~(|t_{1}|^{2}+|t_{2}|^{2}). \label{jIII} \eeq
Equating the expressions in Eqs.~\eqref{jI} and \eqref{jIII}, we find that
\begin{equation} 1 ~-~ |r_{1}|^{2} ~-~ |r_{2}|^{2} ~=~ |t_{1}|^{2} ~+~
|t_{2}|^{2} \label{cons} \end{equation}
in the basis of eigenstates of $\tau^x \si^z$. We have checked numerically
that the computed values of the various probabilities satisfy Eq.~\eqref{cons}.

We will also calculate the spin current of $\vec{\si}$ in the $\hat x$
direction by taking the expectation of the operator $J_{x}\si^{i}$. Choosing
the $\tau^{x} \si^z$ basis as before, we find that
\beq \la J_{x}\si^{y}\ra_{III} ~=~ -\dfrac{v^2 k}{E} ~[(t_{1}^{*}t_{2}
+t_{1}t_{2}^{*})]. \label{jxsy1} \eeq 
We note that 
\beq J_{x}\si^{y} = - v \tau^z. \label{jxsy2} \eeq 
We therefore anticipate that the expectation values of $J_{x}\si^{y}$ will 
have opposite signs on the two surfaces (which correspond to $\tau^z = \pm 1$);
this is a consequence of opposite helicities of the Dirac electrons on these 
surfaces. We will see that this is borne out by the numerical results 
presented below.

\subsection{Differential conductance}
\label{sec3d}

Having chosen the incident waves in the basis of eigenstates of
$\tau^x \si^z$, we can calculate the transmission probabilities
$|t_i|^2$ and transmitted currents $\la J_x \ra$. We can then
calculate the differential conductance $G$ as follows~\cite{deb2}.
If the system has a large width in the $\hat y$ direction given by
$W$, the net current going from the left of the barrier to the right
is given by \beq I ~=~ q W \int \int \frac{dk_x dk_y}{(2\pi)^2} ~\la
J_x \ra, \label{curr} \eeq where $q$ is the charge of the electrons.
We now change variables from $k_x, ~k_y$ to the energy $E = \hbar v
\sqrt{k_x^2 + k_y^2}$ and the angle of incidence $\ta = \tan^{-1}
(k_x/k_y)$ which goes from 0 to $\pi$. If $\mu_L$ and $\mu_R$ denote
the chemical potentials of the left and right leads which are
attached to the system, then $E$ goes from $\mu_R$ to $\mu_L$ in the
integral in Eq.~\eqref{curr}; we are assuming here that $\mu_L >
\mu_R$. The voltage applied in a lead is related to its chemical
potential as $\mu = qV$. In the zero-bias limit, $\mu_L, ~\mu_R
~\to~ \mu$, the differential conductance is given by 
\beq G ~=~ \frac{dI}{dV} ~=~ \frac{q^2 W \mu}{(2\pi v\hbar)^2} ~\int_0^{\pi}
d\ta ~\la J_x \ra. \label{condG} \eeq 
It is convenient to define a quantity $G_0$ which is the maximum possible 
value of $G$ that arises when the transmission probabilities have the maximum 
possible values, $|t_1|^2 = |t_2|^2 = 1$. Equation~\eqref{jIII} then gives 
$\la J_x \ra = 2v \sqrt{1 - \lm^2/\mu^2} \sin \ta$, where we have used the 
relations $E=\mu$ and $vk = \sqrt{E^2 - \lm^2}$. The conductance in this
case is given by
\beq G_0 ~=~ \frac{q^2 W}{v (\pi \hbar)^2} ~\sqrt{\mu^2 ~-~ \lm^2}.
\label{condG0} \eeq
In the figures presented below, we will plot the dimensionless 
ratio $G/G_0$ whose maximum possible value is 1. In the plots, the
conductance will be calculated at a value of the incident electron
energy $E$ which is equal to $\mu$. We will always choose $E$ to lie in the 
range $\lm < E < m$, so that the energy lies in the upper (positive 
energy) band of the surface states but in the gap of the bulk states; hence 
the bulk states will not contribute to the conductance.

\section{$\de$-function barrier}
\label{sec4}

Before studying the more realistic case of barriers with finite widths, it
turns out to be instructive to study the simpler problem of a $\de$-function
barrier. This can be thought of as the limit of a finite width barrier in
which the barrier height $V_0 \to \infty$ and barrier width $L \to 0$, keeping
the product $V_0 L = c$ fixed. We will discover later that many of the results 
obtained for barriers with finite widths can be understood qualitatively by 
considering the problem of $\de$-function barriers.

\subsection{$\de$-function potential barrier ~-~ single surface}
\label{sec4a}

We first consider the case of a single surface of a TI
with a potential barrier of the form $V_0 \de (x)$. For a single surface,
the wave function is a two-component object. Due to the Dirac
nature of the Hamiltonian, we find that a $\de$-function barrier
produces a discontinuity in the wave function. To show this, we consider
\bea
H &=& v (-i\si^{x}\pa_{y} +i \si^{y}\pa_{x}) ~+~ c ~\de (x), \non \\
H \psi &=& E\psi. \label{de} \eea
Following a procedure similar to the one used to study the effect of a
$\de$-function barrier in a Schr\"odinger equation, we integrate
the second equation in Eq.~\eqref{de} through the $\de$-function at
$x=0$. This gives a matching condition at $x=0$ of the form
\beq \psi_{x\to 0^{+}}= e^{i(c/v)\si^{y}}\psi_{x\to 0^{-}}.
\label{cond1} \eeq
For $x < 0$, the wave function has an incident part with amplitude 1
and a reflected part with amplitude $r$; for $x > 0$, the wave
function has a transmitted part with amplitude $t$. We therefore have
\bea \psi_{x\to 0^{-}} &=& \begin{pmatrix}
1 \\ e^{-i\ta}
\end{pmatrix} ~+~
r\begin{pmatrix}
1 \\ e^{i\ta}
\end{pmatrix}, \non \\
\psi_{x\to 0^{+}} &=& t\begin{pmatrix}
1 \\ e^{-i\ta}
\end{pmatrix}. \eea
Equation~\eqref{cond1} then gives
\bea t\begin{pmatrix}
1 \\ e^{-i\ta}
\end{pmatrix} &=& \begin{pmatrix}
\cos(c/v) & \sin(c/v) \\
-\sin(c/v) & \cos(c/v)
\end{pmatrix} \non \\
&& \times ~\bigg[ \begin{pmatrix}
1 \\ e^{-i\ta}
\end{pmatrix} ~+~ r \begin{pmatrix}
1 \\ e^{i\ta}
\end{pmatrix} \bigg]. \eea

The above equation gives two equations involving two variables $r$ and $t$. 
Solving them we obtain
\bea r &=& \dfrac{\sin (c/v) ~(1~+~ e^{-i2\ta})}{2~[i \cos(c/v) \sin \ta ~-~
\sin (c/v) ]}, \non \\
t &=& \dfrac{i\sin\ta}{i\cos (c/v) \sin\ta ~-~ \sin (c/v)}. \eea
It can be verified that $|r|^2 + |t|^{2}$ =1. For $c/v= 2n\pi$, we find that 
$t=1$ and $r=0$, while for $c/v= (2n+1)\pi$, we get $t=-1$ and $r=0$.

\subsection{$\de$-function potential barrier ~-~ two surfaces}
\label{sec4b}

Next we consider the case where we have top and bottom surfaces of a
TI with a coupling $\lm$ between them. We apply a $\de$-function
potential barrier $V_0 \de (x)$ to only the top surface. The wave function 
$\psi$ now has four components with the first two corresponding to the top 
surface and the last two to the bottom. The boundary condition at $x=0$ on the 
top surface remains the same as in the previous section. Using the wave 
vectors obtained in Eq.~\eqref{sim1} and Eq.~\eqref{sim2} in order to obtain 
the condition from Eq.~\eqref{cond1}, we get
\begin{eqnarray} && t_1 |1_{in}\rangle ~+~ t_2 |-1_{in}\rangle \non \\
&& =~ \Lambda_0 ~\left[ |1_{in}\rangle ~+~ r_1 |1_{ref}\rangle ~+~ r_2 
|-1_{ref} \rangle \right], \label{lam} \end{eqnarray}
where the matrix $\Lambda_0$ is given by 
\bea \Lambda_0 &=& \begin{pmatrix} \cos (c/v) & \sin (c/v) & 0 &0 \\ 
-\sin (c/v) & \cos (c/v) & 0 & 0 \\ 
0 & 0 & 1 & 0\\ 0 & 0 & 0 & 1 \end{pmatrix}, \eea
$r_{1}$ and $t_{1}$ are the reflection and transmission amplitudes
for $\ket{1_{in}}$, $r_{2}$ and $t_{2}$ are the reflection and
transmission amplitudes for $\ket{-1_{in}}$, and the incident wave
vector has been chosen to be $\ket{1_{in}}$.

Solving the four-component equation in Eq.~\eqref{lam}, we get,
for $c/v= 2n\pi$,
\beq r_{1}= 0, ~~~r_{2}=0, ~~~t_{1}=1, ~~~\text{and}~~~ t_{2}=0. \eeq
If we choose the incident wave vector to be $\ket{-1_{in}}$, we get
\beq r_{1}= 0, ~~~r_{2}=0, ~~~t_{1}=0, ~~~\text{and}~~~ t_{2}=1. \eeq
For $c/v = (2n+1)\pi$, we obtain from Eq.~\eqref{lam}
\bea r_{2} &=& 0, ~~~~t_{1} ~=~ 0, \non \\
r_{1} &=& \dfrac{\lm e^{-i\ta}}{iE \sin\ta -\lm \cos\ta}, \non \\
t_{2} &=& -~ \dfrac{i \sin\ta \sqrt{E^2-\lm^2}}{iE\sin\ta -\lm \cos\ta}.
\label{refl} \eea
In contrast to the case of a single surface with $c/v$ equal to an odd
multiple of $\pi$, we see that that the reflection amplitude does not
vanish completely. In Eq.~\eqref{refl}, we see that as $\lm \to 0$, $r_{1}
\to 0$ and $t_{2} \to -1$. As $\lm \to E$, $t_{2} \to 0$ and $r_{1} \to -1$.
Similarly for $\ket{-1_{in}}$ as the incident wave vector, we obtain
\bea r_{1} &=& 0, ~~~~t_{2} ~=~ 0, \non \\
r_{2} &=& -~ \dfrac{\lm e^{-i\ta}}{iE\sin\ta ~+~ \lm \cos\ta}, \non \\
t_{1} &=& -~ \dfrac{i\sin\ta \sqrt{E^2-\lm^2}}{iE\sin\ta ~+~ \lm \cos\ta}. \eea

\begin{figure}[H]
\hspace*{-.4cm} \includegraphics[scale=0.5]{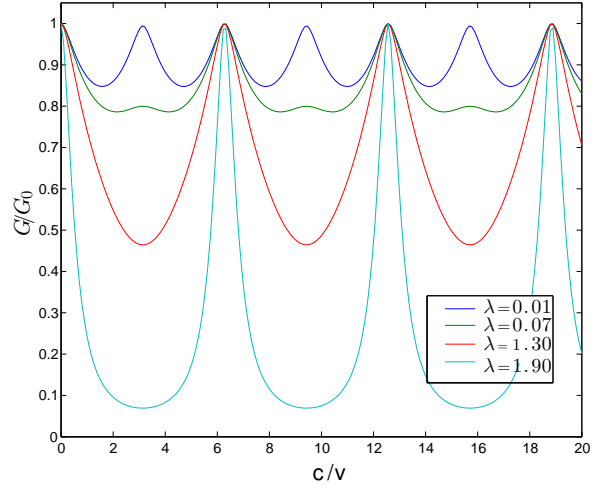} 
\caption{Conductance as a function of $c/v$ for different couplings $\lm$,
when both incident waves are present, and $E=2$.} \label{fig02} \end{figure}

\begin{figure}[H]
\includegraphics[scale=0.6]{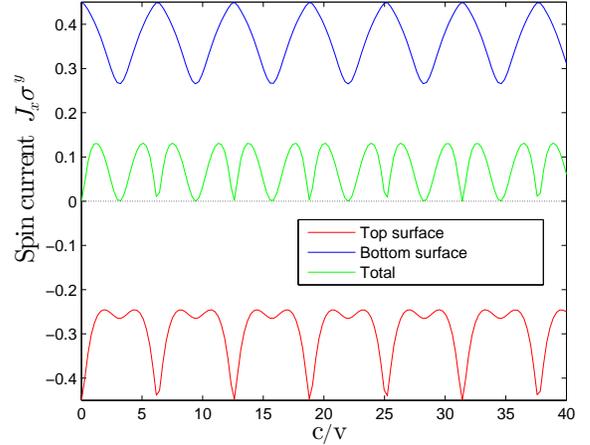} 
\caption{Transmitted spin currents (in units of $v$) as a function of $c/v$ 
when both incident waves are present, $E=2$ and $\lm =1$.} \label{fig03} 
\end{figure}

In Figs.~\ref{fig02} and \ref{fig03}, we show the differential conductance
and spin currents as a function of $c/v$ for certain values of $\lm$; in both
cases, we assume that both incident waves $\ket{1_{in}}$ and $\ket{-1_{in}}$
are present, and we integrate over the angle of incidence $\ta$. 
Figure~\ref{fig02} shows that the oscillation period of the conductance as a 
function of $c/v$ changes from $\pi$ to $2\pi$ as $\lm$ increases; the 
amplitude of the oscillations increases as $\lm$ increases from 0 to $E$.
These observations agree with our
analytic results for a single surface and two coupled surfaces.
In Fig.~\ref{fig03} we show the transmitted spin currents integrated
over $\ta$ at the top and bottom surfaces as a function of $c/v$. As mentioned
above, the spin current at the top (bottom) surface is negative (positive)
although the sum of the two is positive. The oscillation period at both
surfaces is $2\pi$. (In Figs.~\ref{fig02}-\ref{fig05}, the values of $E$ and
$\lm$ are in units of $0.01$ eV, $c/v$ is in units of $\hbar$,
and $\la J_x \si^y \ra$ is in units of $v$).

\subsection{$\de$-function magnetic barrier ~-~ single surface}
\label{sec4c}

Now we consider a $\de$-function magnetic barrier of the form
$V_0 \de (x) \si^{x}$ on the surface of a TI. We have
\bea
H &=& v (-i\si^{x}\pa_{y} + i \si^{y}\pa_{x}) ~+~ c ~\de (x) ~\si^{x}, \non \\
H\psi &=& E\psi. \eea
Integrating over the $\de$-function at $x=0$ now gives
the following matching condition for the wave function,
\beq \psi_{x\to 0^{+}} ~=~ e^{(c/v) \si^z}\psi_{x\to 0^{-}}. \label{cond2} \eeq
(Interestingly, Eqs.~\eqref{cond1} and \eqref{cond2} both satisfy
continuity of the current $\psi^\dagger J_x \psi$ at $x=0$, although
Eq.~\eqref{cond1} is a unitary transformation while Eq.~\eqref{cond2} is not).
Using the same wave functions as in the case of a $\de$-function potential
barrier, we obtain
\beq t\begin{pmatrix}
1 \\ e^{-i\ta}
\end{pmatrix} ~~ = ~~ \begin{pmatrix}
e^{c/v} & 0 \\
0 & e^{-c/v}
\end{pmatrix} \bigg[ \begin{pmatrix}
1 \\ e^{-i\ta}
\end{pmatrix} ~+~ r \begin{pmatrix}
1 \\ e^{i\ta} \end{pmatrix} \bigg]. \eeq
Solving for $r$ and $t$, from the two conditions above, we get
\bea r &=& - ~\dfrac{e^{-2c/v}-1}{e^{-2c/v} e^{i2\ta}-1}, \non \\
t &=& \dfrac{e^{-c/v}(e^{i2\ta}-1)}{e^{-2c/v} e^{i2\ta}-1}. \eea
It can be checked that $|r|^{2} + |t|^{2} = 1$. In the limit $c/v \to \infty$,
we get $r= -1$ and $t=0$. Hence the transmission probability goes to zero as 
the strength of the barrier increases; this is in contract to the 
$\de$-function potential barrier where the transmission probability 
oscillates with the barrier strength.

\subsection{$\de$-function magnetic barrier ~-~ two surfaces}
\label{sec4d}

Similar to the case of a $\de$-function potential barrier, we apply
a $\de$-function magnetic barrier on the top surface of a TI, with
the bottom surface being coupled to the top with the coupling $\lm$ as usual. 
The same boundary condition in this case gives the following equation for 
the case that the incident wave vector is chosen to be $\ket{1_{in}}$,
\begin{eqnarray} && t_1 |1_{in}\rangle ~+~ t_2 |-1_{in}\rangle \non \\
&& =~ \Lambda_1 ~\left[ |1_{in}\rangle ~+~ r_1 |1_{ref}\rangle ~+~ 
r_2 |-1_{ref} \rangle \right], \label{lam2} \end{eqnarray}
where the matrix $\Lambda_1$ is given by 
\bea \Lambda_1 &=& \begin{pmatrix} \exp(c/v) & 0 & 0 &0 \\ 
0 & \exp(-c/v) & 0 & 0 \\ 
0 & 0 & 1 & 0 \\ 
0 & 0 & 0 & 1 \end{pmatrix}. \eea
Here $r_{1}$ and $t_{1}$ are the reflection and transmission
amplitude of $\ket{1_{in}}$, and $r_{2}$ and $t_{2}$ are the
reflection and transmission amplitude of $\ket{-1_{in}}$. Upon
solving these equations in the limit $c/v \to \infty$, we get\\
\bea t_{1} &=& \dfrac{\sin\ta ~(E^{2}-\lm^{2})}{(2E^{2}-\lm^2)\sin\ta ~-~i
\lm^{2}\cos\ta}, \non \\
r_{1} &=& -~ \dfrac{E\sin\ta ~(E+\lm) ~+~ \lm^{2}e^{-i\ta}}{(2E^{2}-\lm^2)
\sin\ta ~-~ i\lm^{2}\cos\ta}, \non \\
r_{2} &=& -~ \dfrac{E\sin\ta ~\sqrt{E^{2}-\lm^{2}}}{(2E^{2}-\lm^2)\sin\ta
~-~ i\lm^{2}\cos\ta}, \non \\
t_{2} &=& -~ \dfrac{\sin\ta(E-\lm) ~\sqrt{E^{2}-\lm^{2}}}{(2E^{2}-\lm^2)
\sin\ta ~-~ i\lm^{2}\cos\ta}. \label{rt} \eea
We see that unless $E\to \lm$, the transmission probability does not vanish
even in the limit of $c/v \to \infty$. This is because the bottom surface
(which does not have a magnetic barrier) allows for the conduction of
electrons since it is coupled to the top surface.

\begin{figure}[H]
\centering
\hspace*{-.4cm} \includegraphics[scale=0.61]{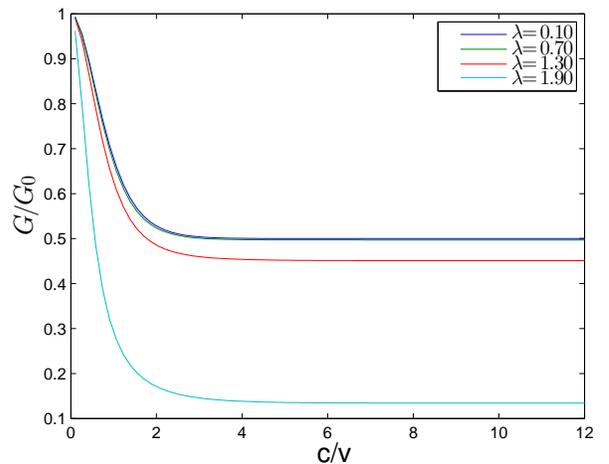} 
\caption{Conductance as a function of $c/v$ for different values of $\lm$,
when both incident waves are present, and $E=2$.} \label{fig04} \end{figure}

\begin{figure}[H]
\centering
\hspace*{-.4cm} \includegraphics[scale=0.61]{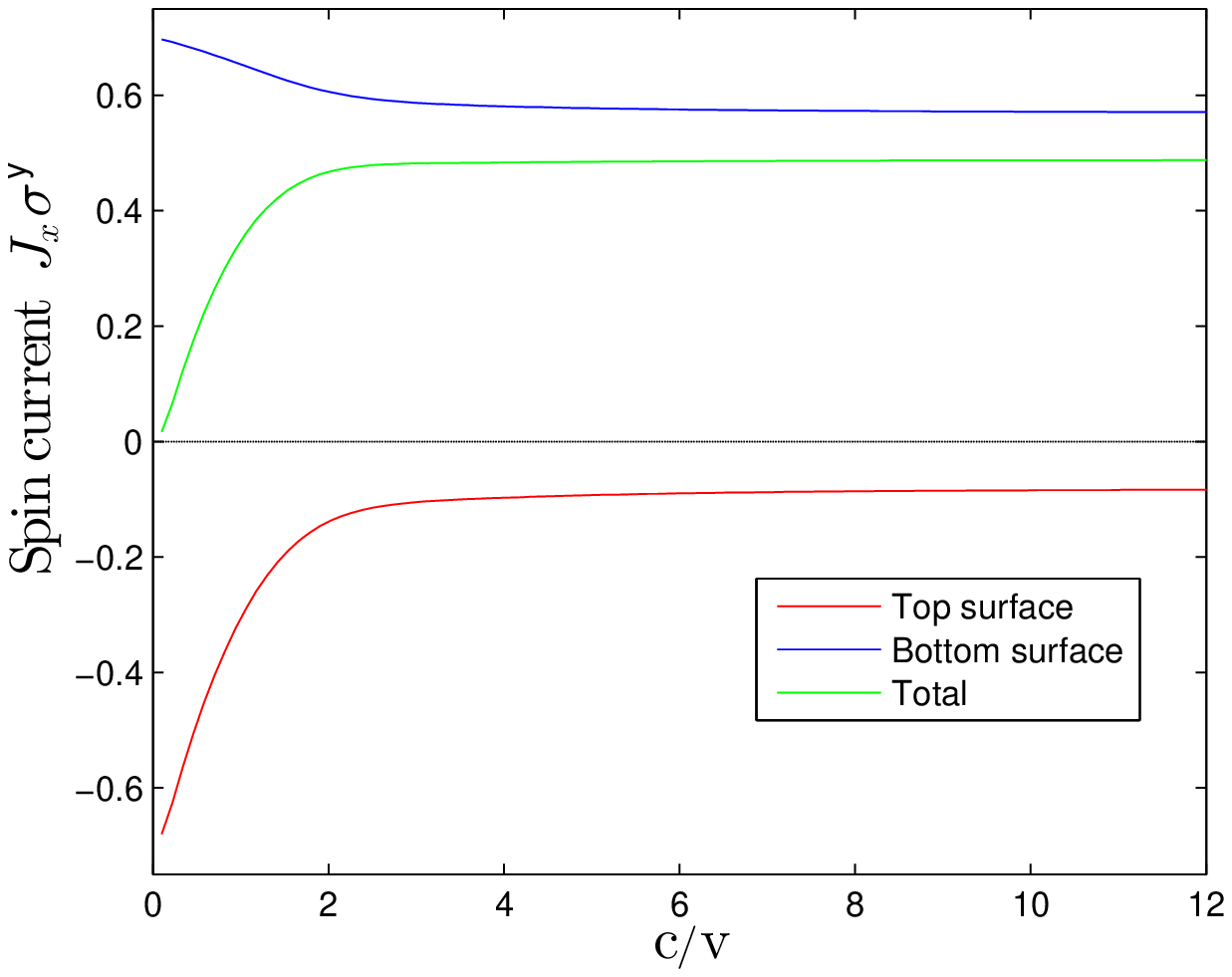} 
\caption{Transmitted spin currents (in units of $v$) integrated over $\ta$ as 
a function of $c/v$, when both incident waves are present, $E=2$ and $\lm =1$.}
\label{fig05} \end{figure}

Figure~\ref{fig04} shows the differential conductance as a function of
$c/v$ for various values of $\lm$; we see that there are no oscillations,
unlike the case of a $\de$-function potential barrier (Fig.~\ref{fig02}).
For a very large value of $c/v$, the conductance does not vanish but reaches
a constant value. However, as $\lm$ approaches $E$, the conductance approaches
zero for a large barrier. This matches with the analytic expressions presented
in Eqs.~\eqref{rt}. Figure~\ref{fig05} shows the transmitted spin current as a
function of $c/v$; this too does not show any oscillations.

\section{Potential barrier with finite width}
\label{sec5}

We now study the case of a finite width potential barrier on the top surface.
In region $II$ where the potential is nonzero, the Hamiltonian is
\begin{equation} H_{II}= \begin{pmatrix}
V_{0} & vk'e^{i\ta'} & \lm & 0 \\
vk'e^{-i\ta'} & V_{0} & 0& \lm \\
\lm & 0 & 0 & -vk'e^{i\ta'}\\
0& \lm & -vk'e^{-i\ta'}& 0
\end{pmatrix} \end{equation}
The eigenvalues of $H_{II}$ and the respective eigenstates are
\bea e_1' &=& \dfrac{V_{0}}{2}+ \dfrac{1}{2}\sqrt{(V_{0}+2vk')^{2}+4 \lm^{2}}
\non \\
{\rm and} ~~~\ket{e_1'} &=& \frac{1}{\sqrt{2(1+\al_1^2)}} ~\begin{pmatrix}
1\\ e^{-i\ta'} \\ \al_{1}\\ \al_{1}e^{-i\ta'} \end{pmatrix}, \non \\
e_{2}' &=& \dfrac{V_{0}}{2}-\dfrac{1}{2}\sqrt{(V_{0}+2vk')^{2}+4\lm^{2}} \non \\
{\rm and} ~~~\ket{e_2'} &=& \frac{1}{\sqrt{2(1+\al_2^2)}} ~\begin{pmatrix}
1\\ e^{-i\ta'}\\ \al_{2}\\ \al_{2}e^{-i\ta'} \end{pmatrix}, \non \\
e_{3}' &=& \dfrac{V_{0}}{2}+ \dfrac{1}{2}\sqrt{(V_{0}-2vk')^{2}+4\lm^{2}} \non \\
{\rm and}~~~ \ket{e_3'} &=& \frac{1}{\sqrt{2(1+\al_3^2)}} ~\begin{pmatrix}
1\\ -e^{-i\ta'}\\ \al_{3}\\ -\al_{3}e^{-i\ta'} \end{pmatrix}, \non \\
e_{4}' &=& \dfrac{V_{0}}{2}- \dfrac{1}{2}\sqrt{(V_{0}-2vk')^{2}+4\lm^{2}} \non \\
{\rm and}~~~ \ket{e_4'} &=& \frac{1}{\sqrt{2(1+\al_4^2)}} ~\begin{pmatrix}
1\\ -e^{-i\ta'}\\ \al_{4}\\ -\al_{4}e^{-i\ta'} \end{pmatrix},
\label{potreg} \eea
where
\bea \al_{1} &=& \dfrac{-V_{0}-2vk'+\sqrt{(V_{0}+2vk')^{2}+4\lm^{2}}}{2\lm},
\non \\
\al_{2} &=& \dfrac{-V_{0}-2vk'-\sqrt{(V_{0}+2vk')^{2}+4\lm^{2}}}{2\lm}, \non \\
\al_{3} &=& \dfrac{-V_{0}+2vk'+\sqrt{(V_{0}-2vk')^{2}+4\lm^{2}}}{2\lm}, \non \\
\al_{4} &=& \dfrac{-V_{0}+2vk'-\sqrt{(V_{0}-2vk')^{2}+4\lm^{2}}}{2\lm}, \non \\
k' e^{i\ta'} &=& k_y+ik_x' ~~~{\rm and}~~~ k^{'2} ~=~ k_x^{'2}+k_y^{2}. \eea
[In the limit $\lm \to 0$, we note that the states labeled
1 and 3 reduce to states at the top surface
(namely, $e_1' \to V_0 + vk'$, $e_3' \to V_0 - vk'$, and the lower
two components of $\ket{e_1'}$ and $\ket{e_3'} \to 0$), while states 2 and 4
reduce to states at the bottom surface (namely, $e_2' \to - vk'$, $e_4' \to vk'$,
and the upper two components of $\ket{e_2'}$ and $\ket{e_4'} \to 0$. We have
assumed here that $V_0 \pm 2 vk' > 0$].
To find the allowed values of $k_x'$, we note that $k_y$ is conserved,
i.e., has the same value as in the barrier-free region, because the
potential is independent of $y$. We now equate the four eigenvalues
shown in Eq.~\eqref{potreg} to the energy $E$ in the barrier-free region
since the energy is conserved. We then obtain for $k'$ the expression
\beq k' ~=~ \pm \frac{1}{2v}(V_{0} \pm \sqrt{(2E-V_{0})^2-4\lm^{2}}), \eeq
in which all four combinations of plus and minus signs can appear.
We then find that
\bea k_{x1}' &=& \sqrt{\dfrac{1}{v^2} \bigg( -\dfrac{V_{0}}{2}+\sqrt{\big(E-
\dfrac{V_0}{2}\big)^{2}-\lm^{2}} \bigg)^{2}-k_y^{2}}~, \non \\
k_{x2}' &=& - ~\sqrt{\dfrac{1}{v^2} \bigg( -\dfrac{V_{0}}{2}+\sqrt{\big(E-
\dfrac{V_0}{2}\big)^{2}-\lm^{2}} \bigg)^{2}-k_y^{2}}~, \non \\
k_{x3}' &=& \sqrt{\dfrac{1}{v^2} \bigg( -\dfrac{V_{0}}{2}-\sqrt{\big(E-
\dfrac{V_0}{2}\big)^{2}-\lm^{2}} \bigg)^{2}-k_y^{2}}~, \non \\
k_{x4}' &=& - ~\sqrt{\dfrac{1}{v^2} \bigg( -\dfrac{V_{0}}{2}-\sqrt{\big(E-
\dfrac{V_0}{2}\big)^{2}-\lm^{2}} \bigg)^{2}-k_y^{2}}~. \non \\
&& \eea

\subsection{Numerical results}
\label{sec5a}

We will now present our results for the transmission probabilities $|t_1|^2$
and $|t_2|^2$, transmitted current $\la J_x \ra$, the differential conductance
$G/G_0$, and the transmitted spin current $\la J_x \si^y \ra$ for various
parameter values. In all the plots, the values of $E$, $\lm$ and $V_0$ are
in units of $0.01$ eV, the barrier width $L$ is in units of
$\hbar v/(0.02$ eV) $~\simeq 17$ nm, and the currents are in units of $v$
(we have taken $v= 0.333$ eV-nm as in Bi$_2$Se$_3$).
We have chosen these units of energy and barrier width as they are
experimentally realistic (see Ref.~\onlinecite{kats} where tunneling through 
barriers in single- and bilayer graphene was studied). Further, we want the 
incident energy $E$ to be much smaller than $m = 0.28$ eV (for Bi$_2$Se$_3$) 
so that the bulk states do not contribute to the conductance.

Figures~\ref{fig06} show the transmitted probabilities $|t_i|^2$ and
currents $\la J_x \ra$ for different choices of the incident waves.
In Fig.~\ref{fig06} (a), where the incident wave has been chosen to
be $\ket{1_{in}}$, we see that $|t_{1}|^{2}$ is symmetric about
$\ta= \pi/2$, whereas $|t_{2}|^{2}$, which is the probability of
$\ket{-1_{in}}$ in region $III$, is asymmetric. Similarly, in
Fig.~\ref{fig06} (b), where the incident wave is $\ket{-1_{in}}$, we
see that $|t_{2}|^{2}$ is symmetric, whereas $|t_{1}|^{2}$, which is
the probability of $\ket{1_{in}}$ in region $III$, is asymmetric
about $\ta= \pi/2$. In Fig.~\ref{fig06} (c), both are symmetric as
the transmitted current gets an equal contribution from the two
waves, $\ket{1_{in}}$ and $\ket{-1_{in}}$; this makes $|t_{1}|^{2}$,
$|t_{2}|^{2}$ and the total current symmetric about
$\ta= \pi/2$. We can understand these symmetries as follows. \\

\noi {\bf $\si^{y}$ symmetry:} The symmetry between
$|t_{1}|^{2}$ and $|t_{2}|^{2}$ at the incident angles $\ta$ and $\pi - \ta$
can be understood by looking at the effect of a unitary transformation by
the operator $\si^{y}$. We observe that \\

\noi (i) ~$\si^{y} H(k_{x}, k_{y})\si^{y} = H(k_{x}, -k_{y})$, where
$H(k_{x}, k_{y})$ is the total Hamiltonian in region $II$ given by
\beq H(k_{x}, k_{y}) ~=~ v \tau^z (\si^x k_{y} - \si^y k_{x}) ~+~
\lm \tau^{x} ~+~ \frac{V_{0}}{2}(1+ \tau^z). \label{sym} \eeq \\

\noi (ii) ~Since $\si^y$ anticommutes with $\si^z$, we have
$\tau^x \si^z \si^y = - \si^y \tau^x \si^z$. Hence $\si^{y}$ changes
the eigenvalue of $\tau^x \si^z$ from $+1$ to $-1$, thus changing
$\ket{1_{in}}$ to $\ket{-1_{in}}$, and vice versa. \\

Using the above results, we can understand why (i) $|t_{1}|^{2}$ in
Fig.~\ref{fig06} (a) and $|t_{2}|^{2}$ in Fig.~\ref{fig06} (b) are related by
$k_{y} \to -k_{y}$, i.e., by $\ta \to \pi - \ta$, and (ii) $|t_{2}|^{2}$ in
Fig.~\ref{fig06} (a) and $|t_{1}|^{2}$ in Fig.~\ref{fig06} (b) are also
related by $\ta \to \pi - \ta$. These symmetries imply that the total
transmission probability, $|t_{1}|^{2} + |t_{2}|^{2}$, when both incident waves
are present, must be symmetric under $\ta \to \pi - \ta$. This is consistent
with Fig.~\ref{fig06} (c).

\begin{widetext}
\begin{center}
\begin{figure}[htb]
\subfigure[]{\includegraphics[scale=0.41]{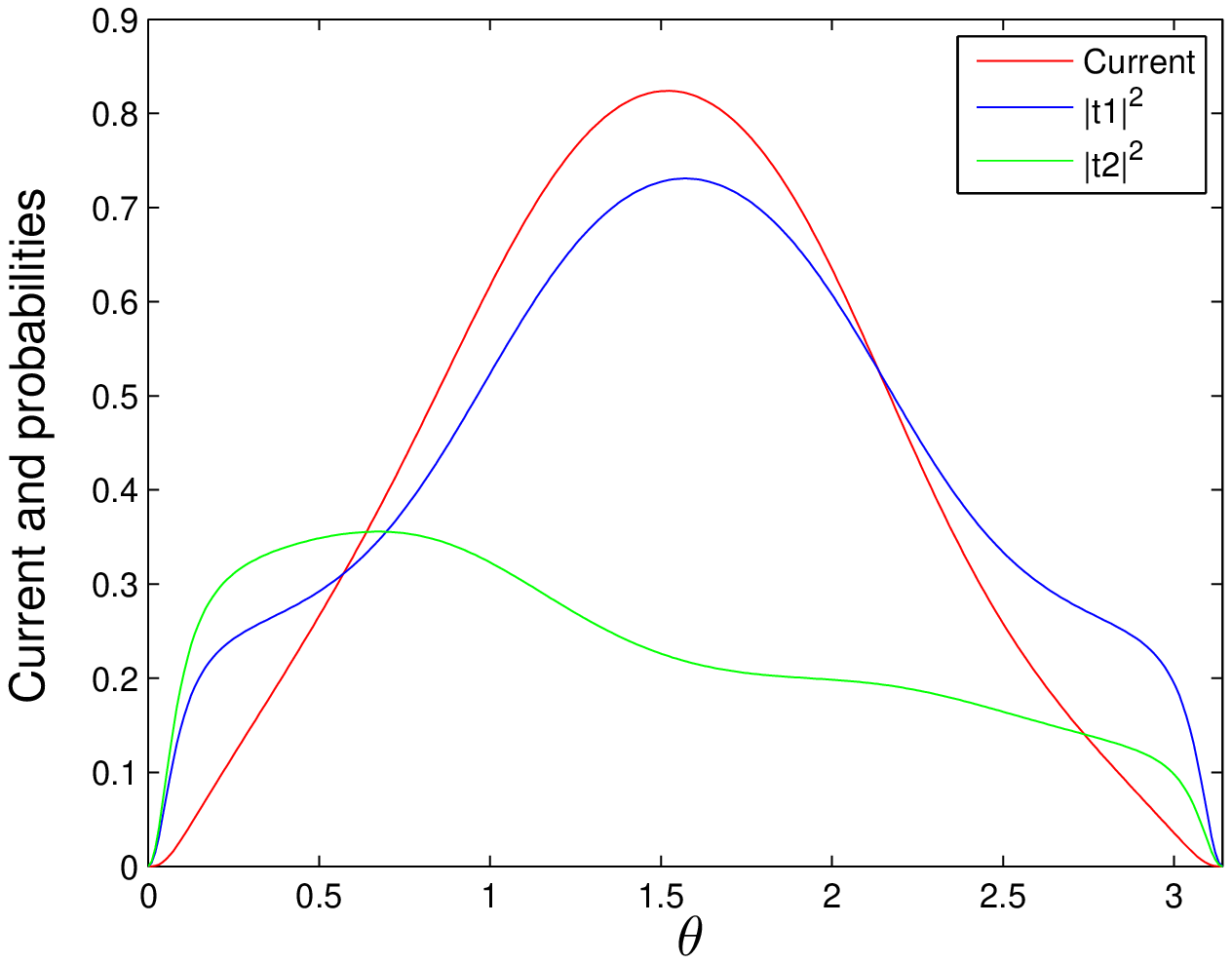}} 
\subfigure[]{\includegraphics[scale=0.41]{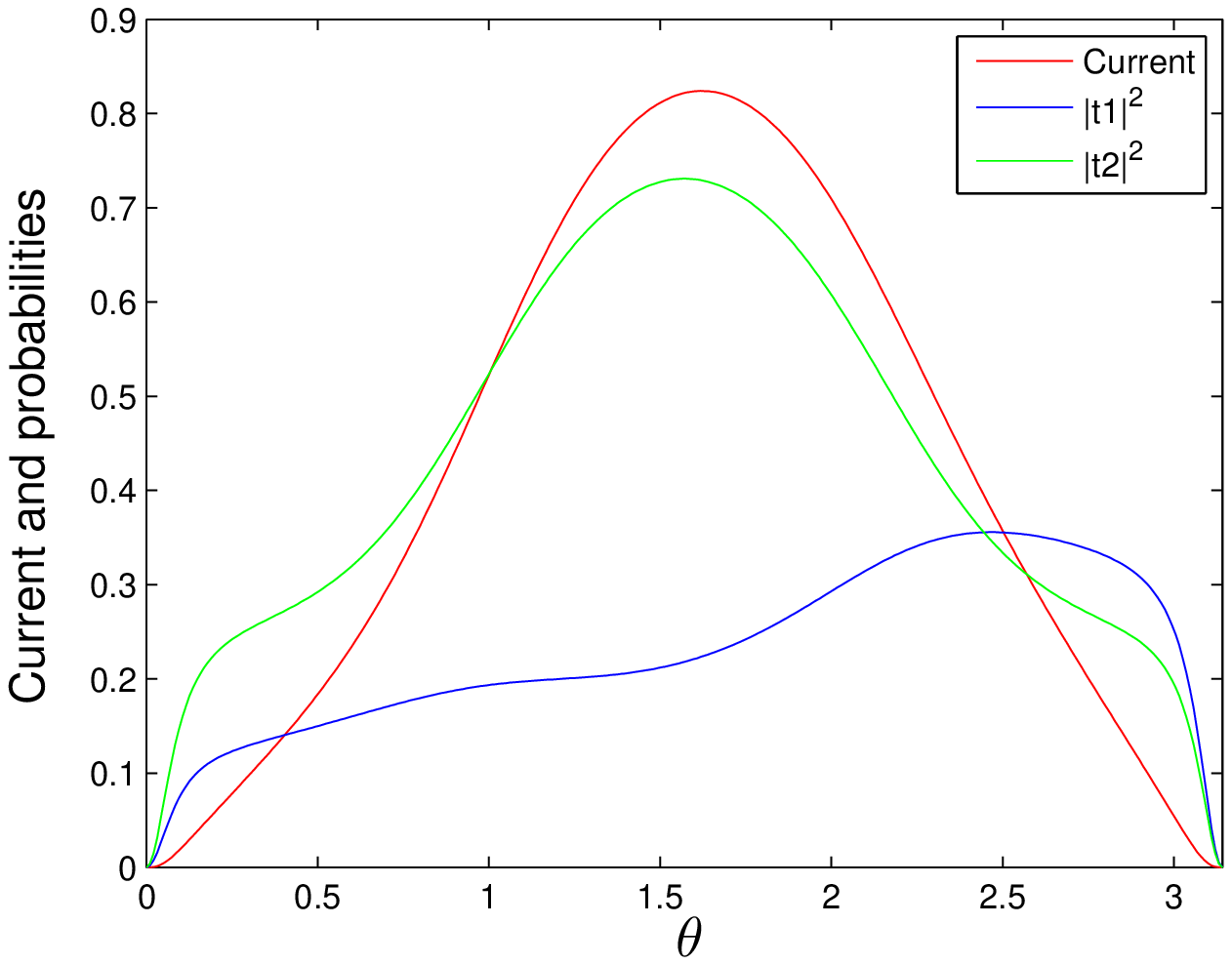}} 
\subfigure[]{\includegraphics[scale=0.41]{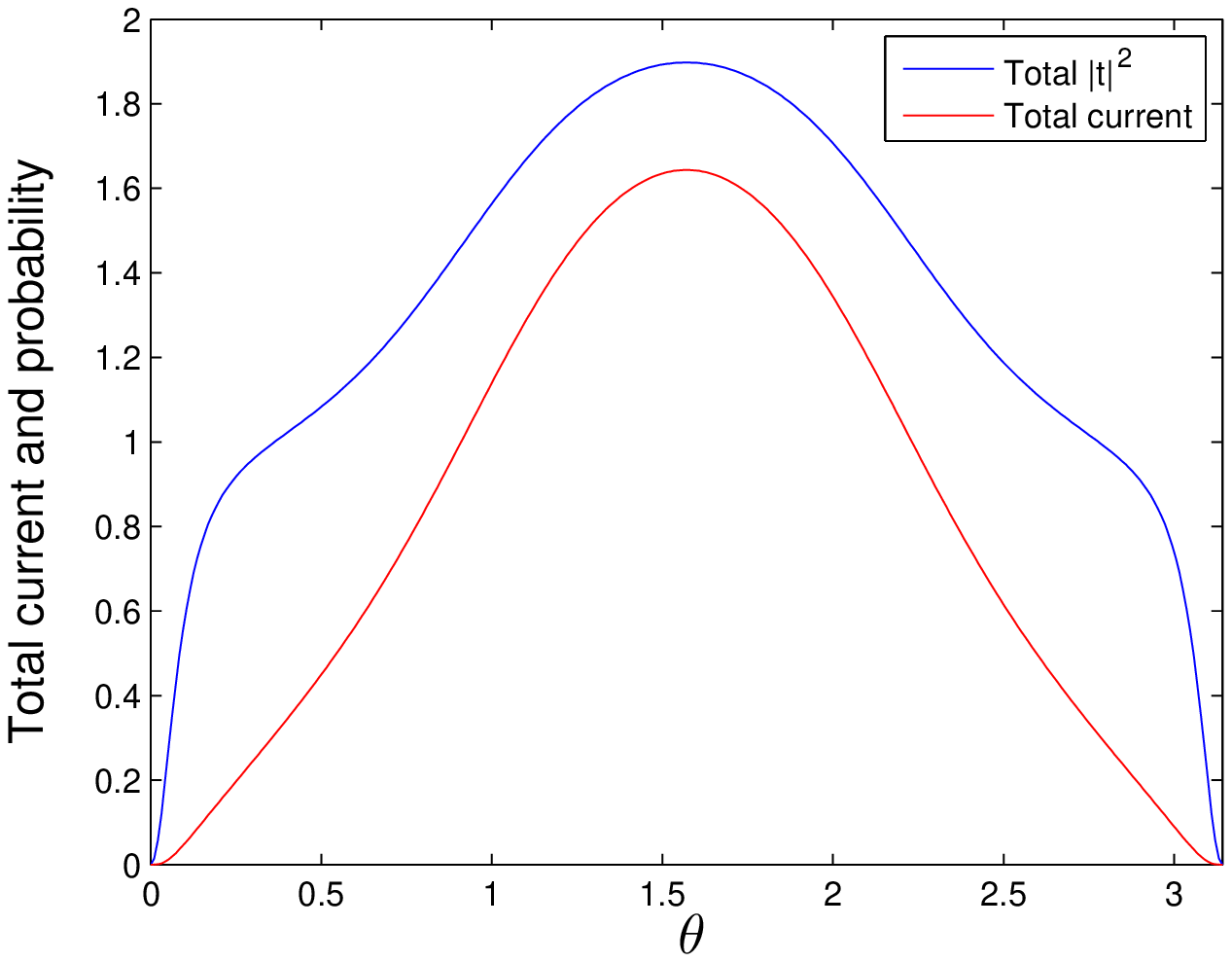}} 
\caption{Transmitted currents (in units of $v$) and probabilities as a function 
of $\ta$ when (a) the incident wave is $\ket{1_{in}}$, (b) the incident wave
function is $\ket{-1_{in}}$, and (c) both incident waves are present. We have
taken $E=2$, $\lm =1$, $V_0 = 1$, and $L=1$.} \label{fig06} 
\end{figure}
\end{center}
\end{widetext}





\begin{figure}[H]
\centering
\hspace*{-.6cm} \includegraphics[scale=0.38]{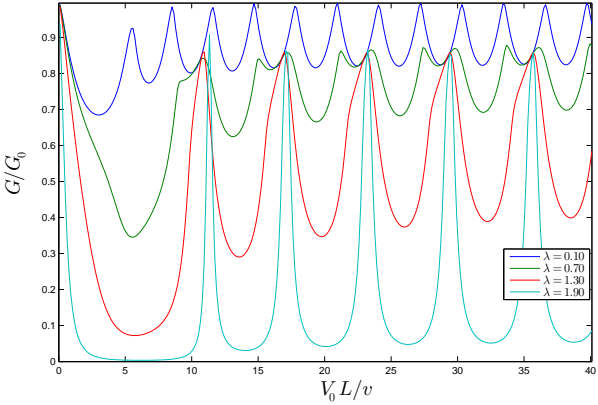} 
\caption{Conductance versus $V_0 L/v$ for different values of $\lm$, with
$E=2$ and $L=1$.} \label{fig07} \end{figure}

In Fig.~\ref{fig07}, the conductance has been plotted as a function of
$V_0 L/v$ (which is in units of $\hbar$). The conductance is seen to oscillate
with a period which depends on the parameter $\lm$. The conductance decreases
with increase in $\lm$ as expected. An interesting phenomenon is that for small
values of $\lm$ (much smaller than the incident energy $E$), the period of
oscillation of the current with $V_0 L/v$ is $\pi$. However, for large values 
of $\lm$, comparable to $E$, the oscillation period is $2\pi$, which is twice 
the previous value. The oscillation period for small $\lm$ can been understood
analytically as follows. For $\lm \simeq 0$, the top and bottom states are 
decoupled, and we can find the transmitted amplitudes analytically. 
For $V_0 \gg E$, we find that
\beq |t_{1}|^{2} ~=~ \dfrac{\sin^{2}\ta}{\sin^2 (V_{0}L/v) \cos^{2}\ta
~+~ \sin^{2}\ta}. \eeq
It is clear that the maxima of $|t_{1}|^{2}$ lie at $V_{0}L/v= n\pi$.
For $\lm$ close to $E$, we do not have an analytical expression for
$|t_{1}|^{2}$, and we therefore do not have an analytical understanding of
the oscillation period. However, we have gained some understanding of this
by looking at the limit of a $\de$-function potential barrier in
Sec.~\ref{sec4}.

\subsection{Currents at the top and bottom surfaces}
\label{sec5b}

It is interesting to look at the currents at the top and bottom surfaces
separately. (It may be possible to experimentally detect these currents
by attaching leads to the system which couple differently to the top
and bottom surfaces). This is done by taking the projections of the previously
obtained transmitted wave functions on to the top and bottom surfaces
(i.e., taking the upper and lower two components, respectively) and
then calculating the expectation value of $J_{x}$ for these wave
functions. (In Eq.~\eqref{jxy}, we note that $J_{x}$ is block diagonal
in the basis of top and bottom surface states).

In region $III$, we have
\bea \ket{1_{in}} &=& \dfrac{1}{2 \sqrt{E}} \begin{pmatrix}
\sqrt{E+\lm} \\ \sqrt{E-\lm}e^{-i\ta} \\ \sqrt{E+\lm} \\ -\sqrt{E-\lm}e^{-i\ta}
\end{pmatrix}, \non \\
\ket{-1_{in}} &=& \dfrac{1}{2 \sqrt{E}} \begin{pmatrix}
\sqrt{E-\lm} \\ \sqrt{E+\lm}e^{-i\ta} \\ -\sqrt{E-\lm} \\ \sqrt{E+\lm}e^{-i\ta}
\end{pmatrix}. \label{sim3} \eea

For the top and bottom surfaces,
\beq \ket{\psi_{III,t/b}} ~=~ (t_{1} \ket{1_{in, t/b}} ~+~ t_{2}
\ket{-1_{in, t/b}}) ~e^{i(k_{x}x+k_{y}y)}, \eeq
where $\ket{1_{in, t/b}} = [(1 \pm \tau^z)/2] \ket{1_{in}}$ and
$\ket{-1_{in, t/b}} = [(1 \pm \tau^z)/2] \ket{-1_{in}}$. Namely,
\bea \ket{1_{in,t}} &=& \dfrac{1}{2 \sqrt{E}} \begin{pmatrix}
\sqrt{E+\lm} \\ \sqrt{E-\lm}e^{-i\ta} \\ 0 \\ 0
\end{pmatrix}, \non \\
\ket{-1_{in,t}} &=& \dfrac{1}{2 \sqrt{E}} \begin{pmatrix}
\sqrt{E-\lm} \\ \sqrt{E+\lm}e^{-i\ta} \\ 0 \\ 0
\end{pmatrix}, \label{sim4} \eea
are the wave functions at the top surface, and
\bea \ket{1_{in,b}} &=& \dfrac{1}{2 \sqrt{E}} \begin{pmatrix}
0\\ 0 \\ \sqrt{E+\lm} \\ -\sqrt{E-\lm}e^{-i\ta}
\end{pmatrix}, \non \\
\ket{-1_{in,b}} &=& \dfrac{1}{2 \sqrt{E}} \begin{pmatrix}
0 \\ 0 \\ -\sqrt{E-\lm} \\ \sqrt{E+\lm}e^{-i\ta} \end{pmatrix},
\label{sim5} \eea
are the wave functions at the bottom surface. Since $J_{x}$ is block diagonal 
in this basis, we can calculate $\la J_{x ,t/b} \ra$ where
\beq J_{x ,t} ~=~ \frac{1 ~+~ \tau^z}{2} ~J_x ~~~~{\rm and}~~~~ J_{x,b} ~=~
\frac{1 ~-~ \tau^z}{2} ~J_x. \eeq
We then get for the top and bottom surfaces
\bea && \bra{\psi_{III,t}} J_{x ,t}\ket{\psi_{III,t}} ~=~ \dfrac{v^2 k}{2E}
\sin\ta ~(|t_{1}|^{2}+|t_{2}|^{2}) \non \\
&+& \dfrac{v}{2E}~[E \sin\ta (t_{1}^{*}t_{2}+t_{1}t_{2}^{*}) ~+~
i\lm \cos\ta(t_{1}^{*}t_{2}-t_{1}t_{2}^{*})], \non \\
&& \bra{\psi_{III,b}} J_{x ,b}\ket{\psi_{III,b}} ~=~ \dfrac{v^2 k}{2E}\sin\ta
~(|t_{1}|^{2}+|t_{2}|^{2}) \non \\
&-& \dfrac{v}{2E}[E\sin\ta (t_{1}^{*}t_{2}+t_{1}t_{2}^{*}) ~+~
i\lm \cos\ta(t_{1}^{*}t_{2}-t_{1}t_{2}^{*})]. \non \\
&& \label{tb1} \eea
Note that the cross-terms do not vanish when we calculate the currents at
the top and bottom surfaces separately, and these terms appear with opposite
signs at the two surfaces. 

Note that when $V_0 = 0$, i.e., there is no scattering, we have no
cross-terms since either $t_1$ or $t_2$ vanishes depending on whether the
incident wave is $\ket{1_{in}}$ or $\ket{-1_{in}}$. We then get equal currents
at the top and bottom surfaces,
\beq \bra{\psi_{III,t}} J_{x ,t}\ket{\psi_{III,t}} ~=~
\bra{\psi_{III,b}} J_{x ,b}\ket{\psi_{III,b}}. \eeq
The difference between the currents at the two surfaces is therefore a 
measure of the barrier strength $V_0$.

Using the expressions in Eq.~\eqref{tb1}, we obtain the results shown in 
Figs.~\ref{fig08} for the currents at the top and bottom surfaces as a 
function of $\ta$. Interestingly, these figures show that the currents
at the top and bottom surfaces separately can have negative values
for certain ranges of $\ta$ when only one one of the incident waves is
present. This means that some current flows from the top surface to the
bottom surface or vice versa. (Typically this happens close to a glancing
angle of incidence, i.e., $\ta \gtrsim 0$ and $\ta \lesssim \pi$). However
the total current when both incident waves are present is positive for
all values of $\ta$ at both surfaces; we can see this in Fig.~\ref{fig08} (c).
We also note that the individual currents are not symmetric about $\ta= \pi/2$
(normal incidence) although the total current is symmetric about $\ta= \pi/2$.

\begin{widetext}
\begin{center}
\begin{figure}[htb]
\subfigure[]{\includegraphics[scale=0.41]{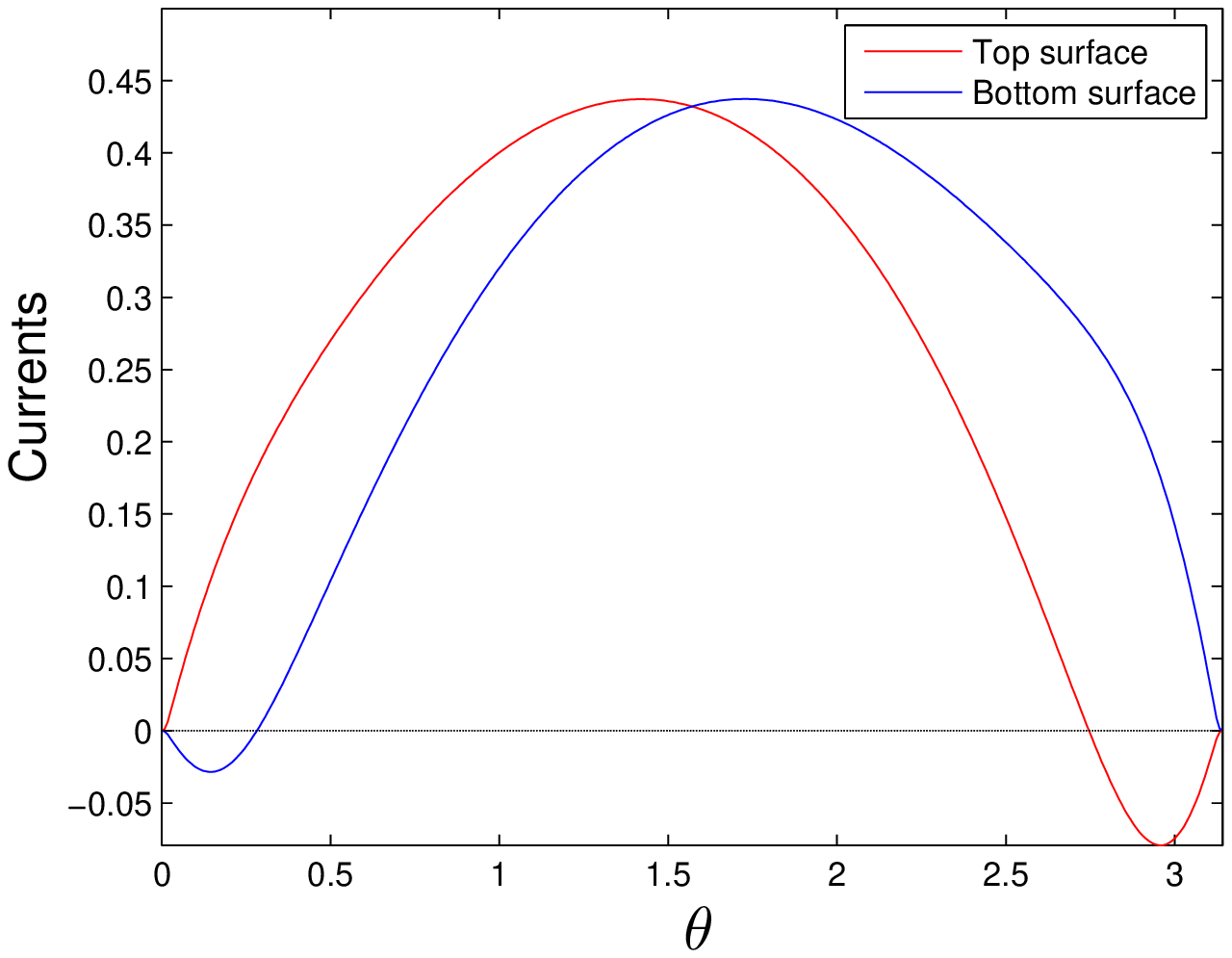}} 
\subfigure[]{\includegraphics[scale=0.41]{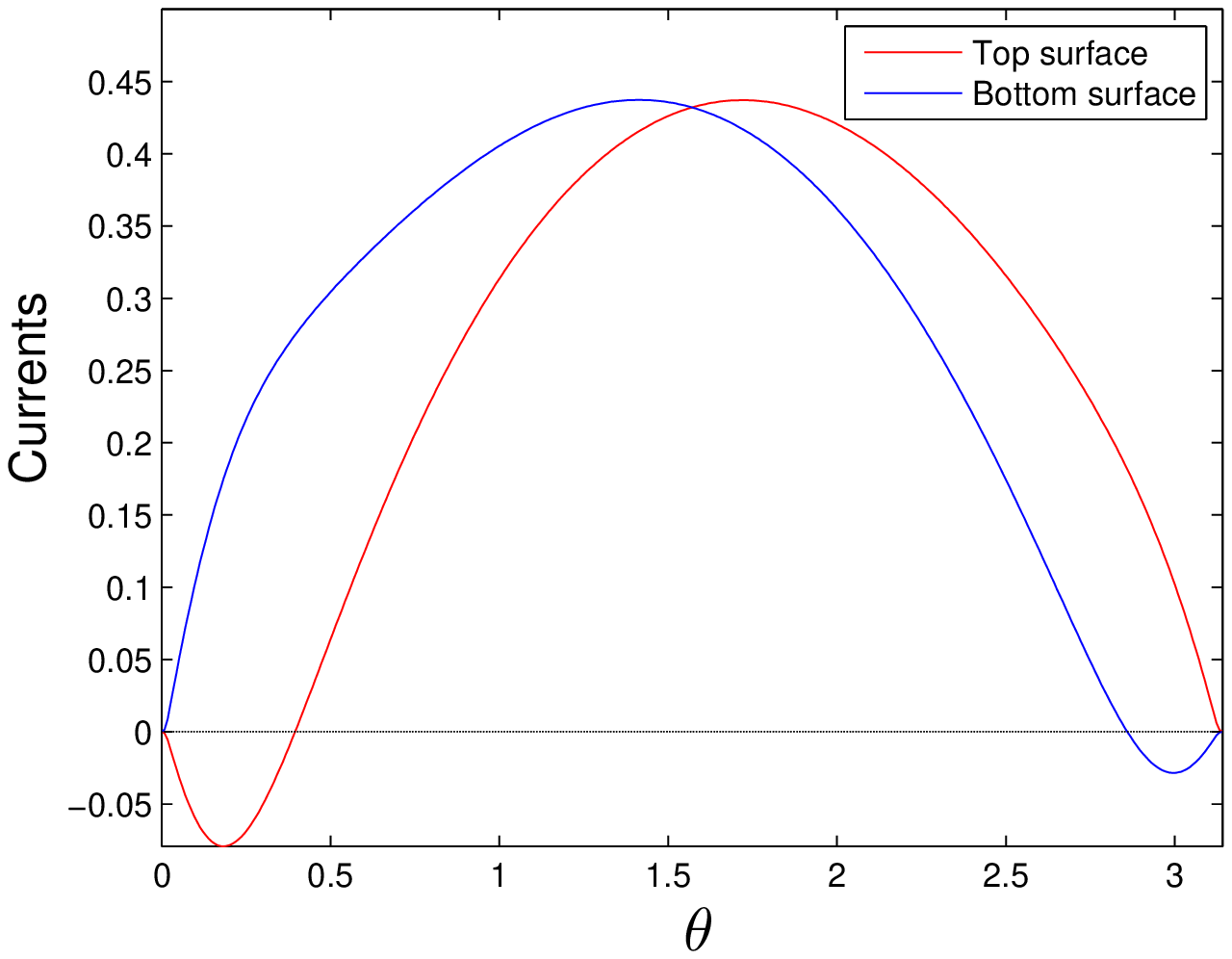}} 
\subfigure[]{\includegraphics[scale=0.41]{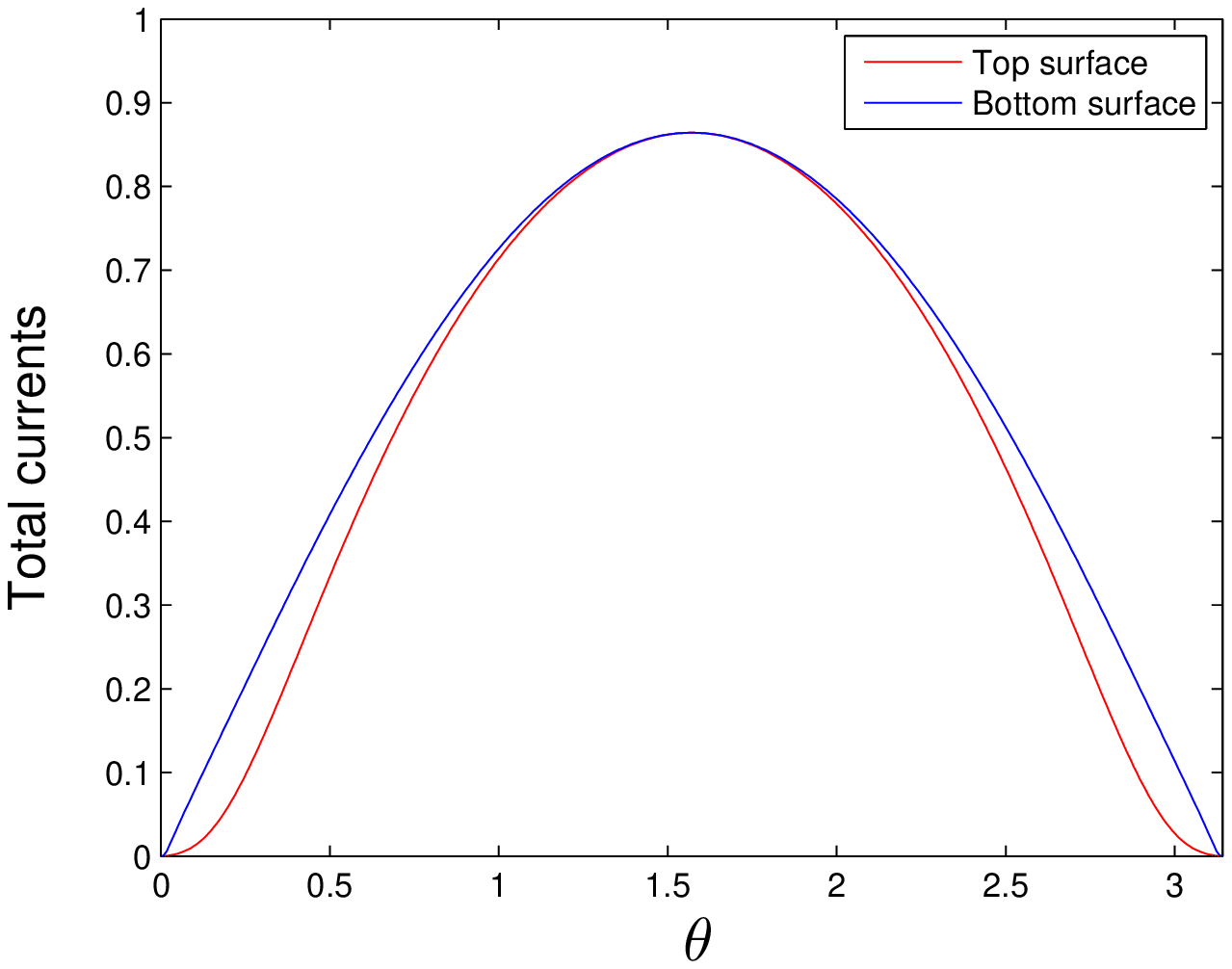}} 
\caption{Transmitted currents (in units of $v$) at the top and bottom surfaces 
as a function of $\ta$ when (a) the incident wave is $\ket{1_{in}}$, (b) the 
incident wave function is $\ket{-1_{in}}$, and (c) both incident waves are 
present. We have taken $E=2$, $\lm =1$, $V_0 = 0.25$, and $L=1$.} \label{fig08}
\end{figure}
\end{center}
\end{widetext}

Figures~\ref{fig09} (a), \ref{fig09} (b) and \ref{fig09} (c) show how the 
conductances at the top and
bottom surfaces vary with $V_0 L/v$. For small values of the coupling $\lm$,
the bottom surface conducts independently of the top surface and gives a
constant current, while the current at the top surface oscillates with a
period $\pi$. As we increase $\lm$, the current at the bottom surface also
begins to develop an oscillatory behavior. Finally, when $\lm$ is close to
$E$, there are sharp peaks which occur with a period equal to $2\pi$. The
variation of the period as $\lm$ increases from zero to $E$ is similar to
the results that we found for a $\de$-function potential barrier in
Sec.~\ref{sec4b}.

\begin{widetext}
\begin{center}
\begin{figure}[htb]
\subfigure[]{\includegraphics[scale=0.41]{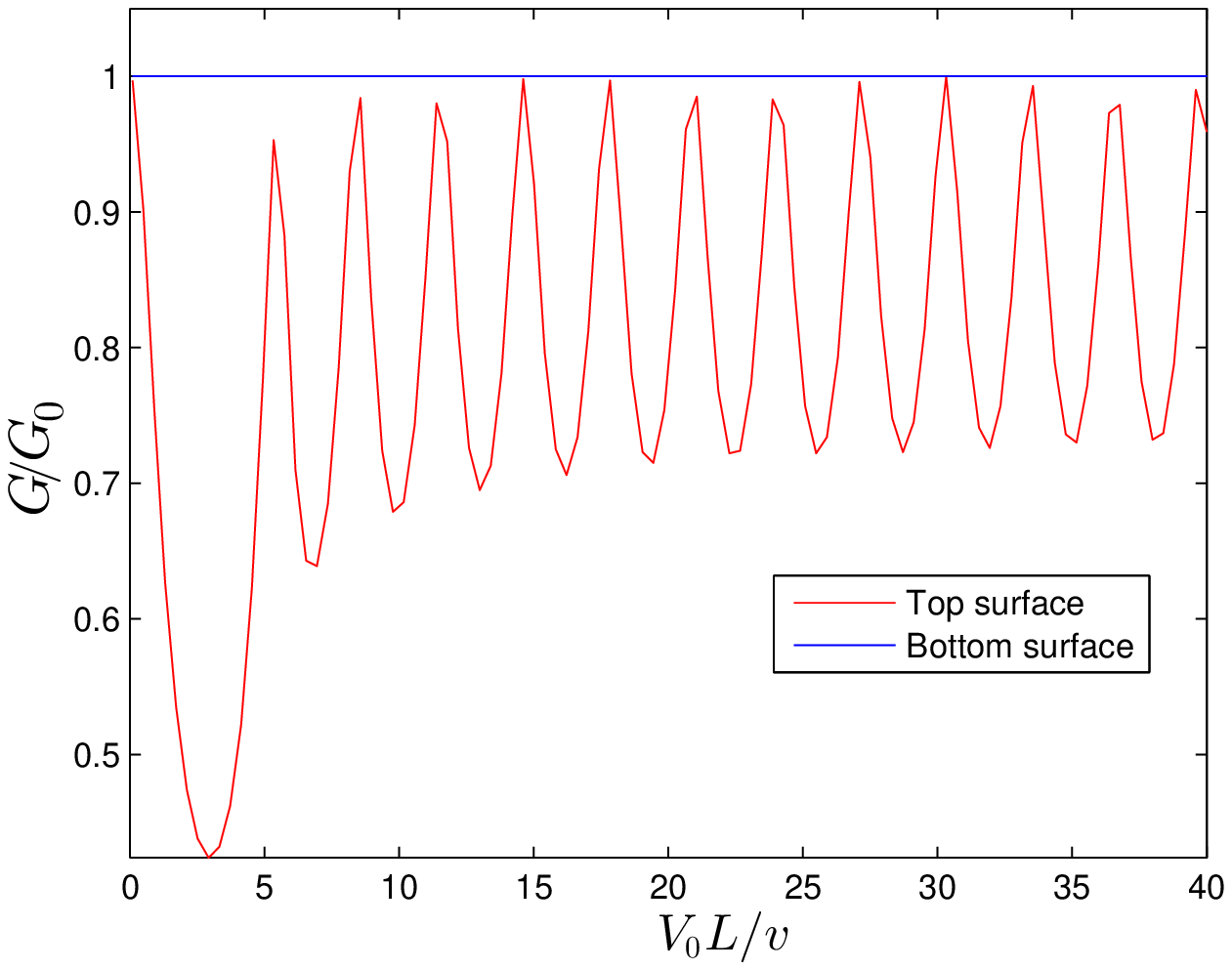}} 
\subfigure[]{\includegraphics[scale=0.41]{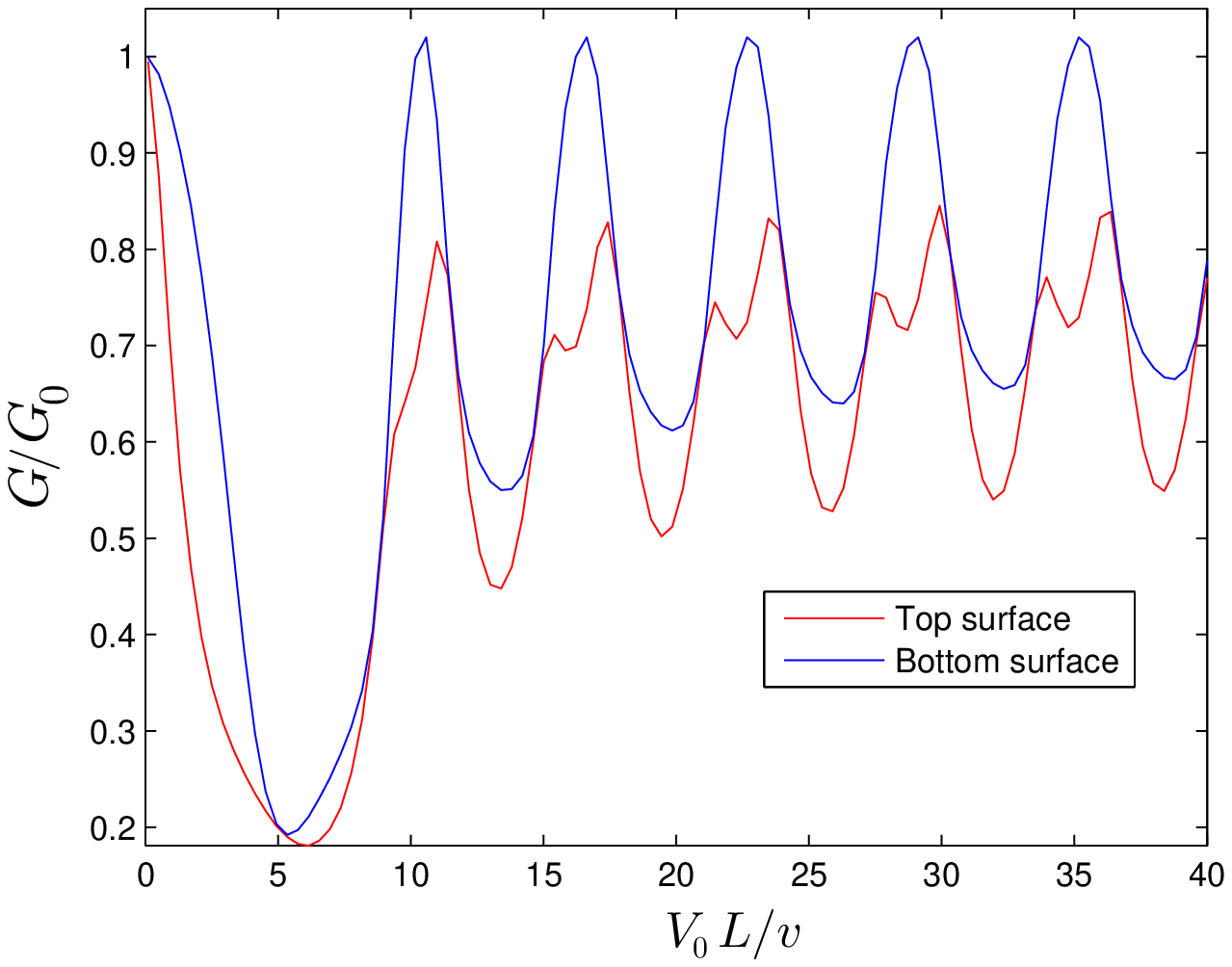}} 
\subfigure[]{\includegraphics[scale=0.41]{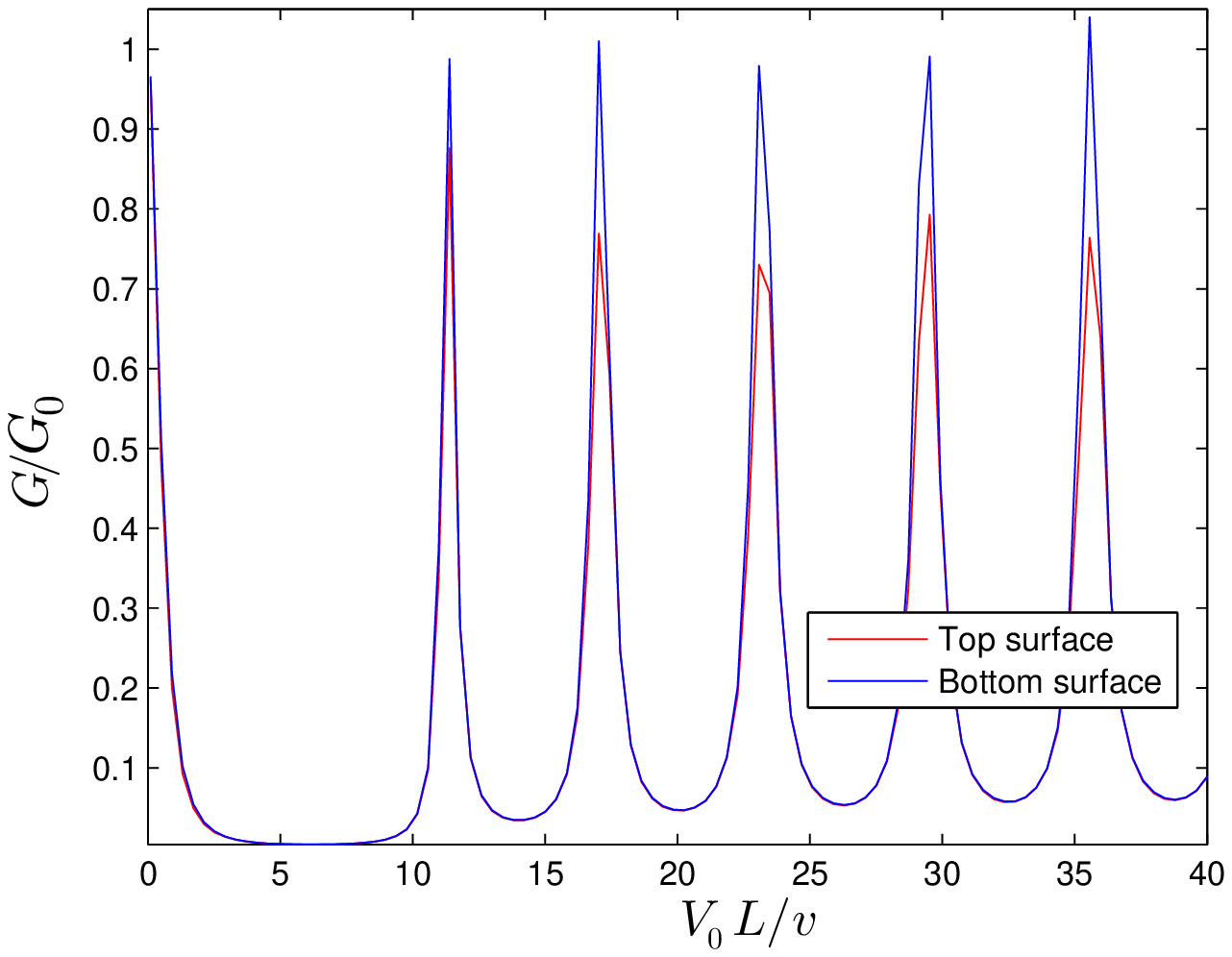}} 
\caption{Conductances at the top and bottom surfaces versus $V_0 L/v$ for (a) 
a small value of $\lm=0.01$, (b) an intermediate value of $\lm=1$, and (c) a 
value of $\lm=1.9$ close to $E$. We have taken $E=2$ and $L=1$.} \label{fig09} 
\end{figure}
\end{center}
\end{widetext}


Similarly, we can obtain the expressions for the spin current ($J_{x} \si^{y}$)
as discussed in Eqs.~\eqref{jxsy1} and \eqref{jxsy2}. For the top and bottom 
surfaces separately, we have to calculate the expectation values of $-(v/2)(1 
+ \tau^z)$ and $(v/2)(1-\tau^z)$, respectively; this gives
\bea \la J_{x}\si^{y}\ra_{t} &=& -~\dfrac{v}{2} ~\big[|t_{1}|^{2}+|t_{2}|^{2}+
\dfrac{vk}{E}(t_{1}^{*}t_{2}+t_{1}t_{2}^{*})\big], \non \\
\la J_{x}\si^{y}\ra_{b} &=& \dfrac{v}{2} ~\big[|t_{1}|^{2}+|t_{2}|^{2}-
\dfrac{vk}{E}(t_{1}^{*}t_{2}+t_{1}t_{2}^{*})\big]. \label{tb2} \eea
In contrast to Eqs.~\eqref{tb1} for the currents at the top and bottom
surfaces, we see that the cross-terms for the spin current appear with the 
same sign at the two surfaces. For $V_0 = 0$, there are no cross-terms and 
the spin currents at the top and bottom surfaces have opposite values,
\beq \la J_{x}\si^{y}\ra_{t} ~=~ -~ \la J_{x}\si^{y}\ra_{b}. \eeq
The total spin current is then zero. Hence the total spin current is a measure
of the barrier strength $V_0$.


\begin{figure}[H]
\centering
\hspace*{-.4cm} \includegraphics[scale=0.62]{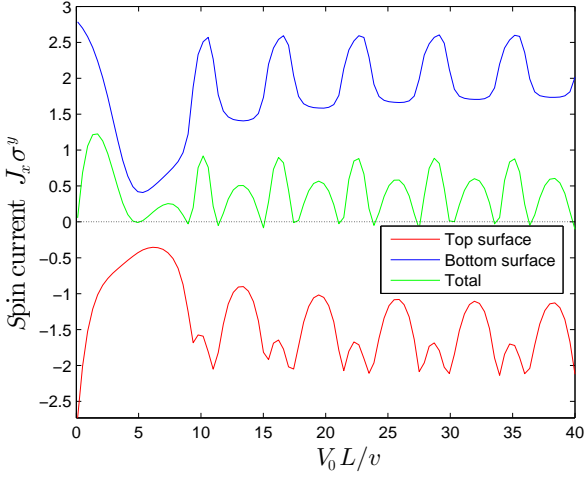} 
\caption{Total spin currents (in units of $v$) as a function of $V_0 L/v$ 
when both incident waves are present, for $E=2$, $\lm =1$, and $L=1$.} 
\label{fig10} \end{figure}

In Fig.~\ref{fig10}, we show the total spin currents as a function of $V_0 L/v$
when both incident waves are present. Once again we see oscillations, the
period of the largest oscillations being $2\pi$. Note also that the spin
current is always negative (positive) at the top (bottom) surface as was
mentioned after Eq.~\eqref{jxsy2}.

\section{Magnetic barrier with finite width}
\label{sec6}

In Sec.~\ref{sec5}, we have studied the effects of a potential barrier with
strength $V_{0}$ on the top surface. We will now study what happens if we
replace the potential barrier by a magnetic barrier of the form $V_{0}\si^{x}$.
This may be experimentally realized by placing a strip of
a ferromagnetic material on the top surface whose magnetization points
along the $\hat x$ direction and has a Zeeman coupling to the spin of the
surface electrons. (For convenience, we will include both the magnetization
of the ferromagnetic strip and its coupling to the electron spin in the
definition of $V_0$ so that it has dimensions of energy).
The Hamiltonian in the barrier region $II$ is now given by
\begin{equation} H_{II}= \begin{pmatrix}
0 & vk' e^{i\ta'}+V_{0} & \lm & 0 \\
vk' e^{-i\ta'} +V_{0}& 0 & 0& \lm \\
\lm & 0 & 0 & -vk' e^{i\ta'}\\
0& \lm & -vk' e^{-i\ta'}& 0 \end{pmatrix}. \end{equation} \\
The eigenvalues of this are found to be
\bea e_{1}' &=& \sqrt{E^2 +V_{0}vk'\cos\ta'+ \frac{V_{0}^{2}}{2} + V_{0} ~A}~,
\non \\
e_{2}' &=& -~\sqrt{E^2 +V_{0}vk'\cos\ta'+ \frac{V_{0}^{2}}{2}+ V_{0} ~A}~,
\non \\
e_{3}' &=& \sqrt{E^2 +V_{0}vk'\cos\ta'+ \frac{V_{0}^{2}}{2} - V_{0} ~A}~,
\non \\
e_{4}' &=& -~\sqrt{E^2 +V_{0}vk'\cos\ta'+ \frac{V_{0}^{2}}{2} -V_{0} ~A}~,
\non \\
A &=& \sqrt{\frac{V_{0}^{2}}{4}+\lm^{2}+V_{0}vk'\cos\ta' + v^{2}k^{'2}
\cos^{2}\ta'}. \label{eig1} \eea
Since the energy $E = \sqrt{v^2 k^2 + \lm^2}$ and $k_{y} = k \cos \ta
= k' \cos \ta'$ are
conserved in all the regions, we find that $k_{x}'$ can take one of the
following values in region $II$,
\bea k_{x1}' &=& \dfrac{1}{v}\sqrt{v^{2} k^{'2} \sin^2 \ta' - V_0 v k' \cos \ta' -
\frac{V_{0}^{2}}{2}+V_{0} ~A}~, \non \\
k_{x2}' &=& -~\dfrac{1}{v}\sqrt{v^{2} k^{'2} \sin^2 \ta' - V_0 v k' \cos \ta' -
\frac{V_{0}^{2}}{2}+V_{0} ~A}~, \non \\
k_{x3}' &=& \dfrac{1}{v}\sqrt{v^{2}k^{'2} \sin^2 \ta' - V_0 v k' \cos \ta' -
\frac{V_{0}^{2}}{2}-V_{0} ~A}~, \non \\
k_{x4}' &=& -~\dfrac{1}{v}\sqrt{v^{2} k^{'2} \sin^2 \ta' - V_0 v k' \cos \ta' -
\frac{V_{0}^{2}}{2}-V_{0} ~A}~, \non \\
&& \eea
where $A$ is defined in Eq.~\eqref{eig1}. \\

\subsection{Numerical results}
\label{sec6a}

Just as for the case of a potential barrier, we will now study how the
current varies with different parameters like the angle of incidence $\ta$,
the coupling $\lm$, and the strength of the magnetic barrier $V_0$.
We present our numerical results below.

\begin{figure}[H]
\centering
\hspace*{-.4cm} \includegraphics[scale=0.62]{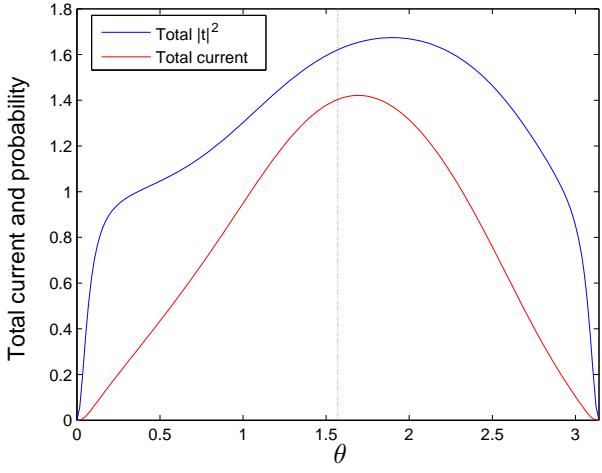} 
\caption{Total transmitted current (in units of $v$) and probability as a 
function of $\ta$ when both incident waves are present. We have taken $E=2$, 
$\lm =1$, $V_0 = 1$, and $L=1$.} \label{fig11} \end{figure}

Figure~\ref{fig11} shows the total transmitted current and probability as a 
function of $\ta$ when both incident waves are present. We see that these are
not symmetric about $\ta= \pi/2 $. This is because the magnetic barrier term
$V_0 \si^x$ breaks the $\si^{y}$ symmetry of the Hamiltonian, unlike the
case of the potential barrier discussed in Eq.~\eqref{sym}.

Figures~\ref{fig12} (a), \ref{fig12} (b) and \ref{fig12} (c) show the 
conductance as a function of
$V_0 L/v$ for different values of $\lm$ and $L$. While we do not see
appreciable oscillations in the conductance if $L$ and $\lm$ are small, more
and more oscillations become visible when $L$ becomes large and $\lm$
approaches $E$.

\begin{widetext}
\begin{center}
\begin{figure}[htb]
\subfigure[]{\includegraphics[scale=0.4]{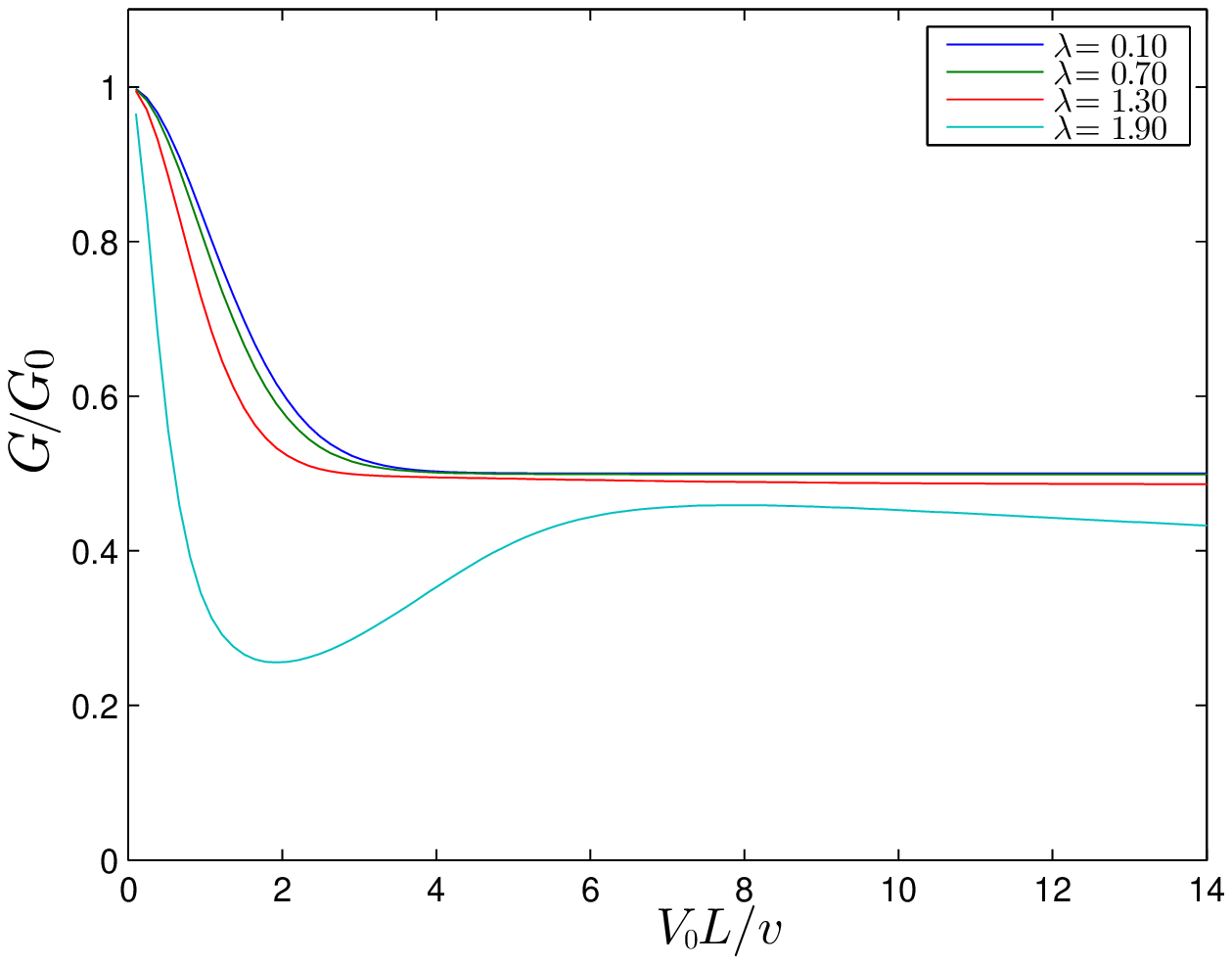}} 
\subfigure[]{\includegraphics[scale=0.43]{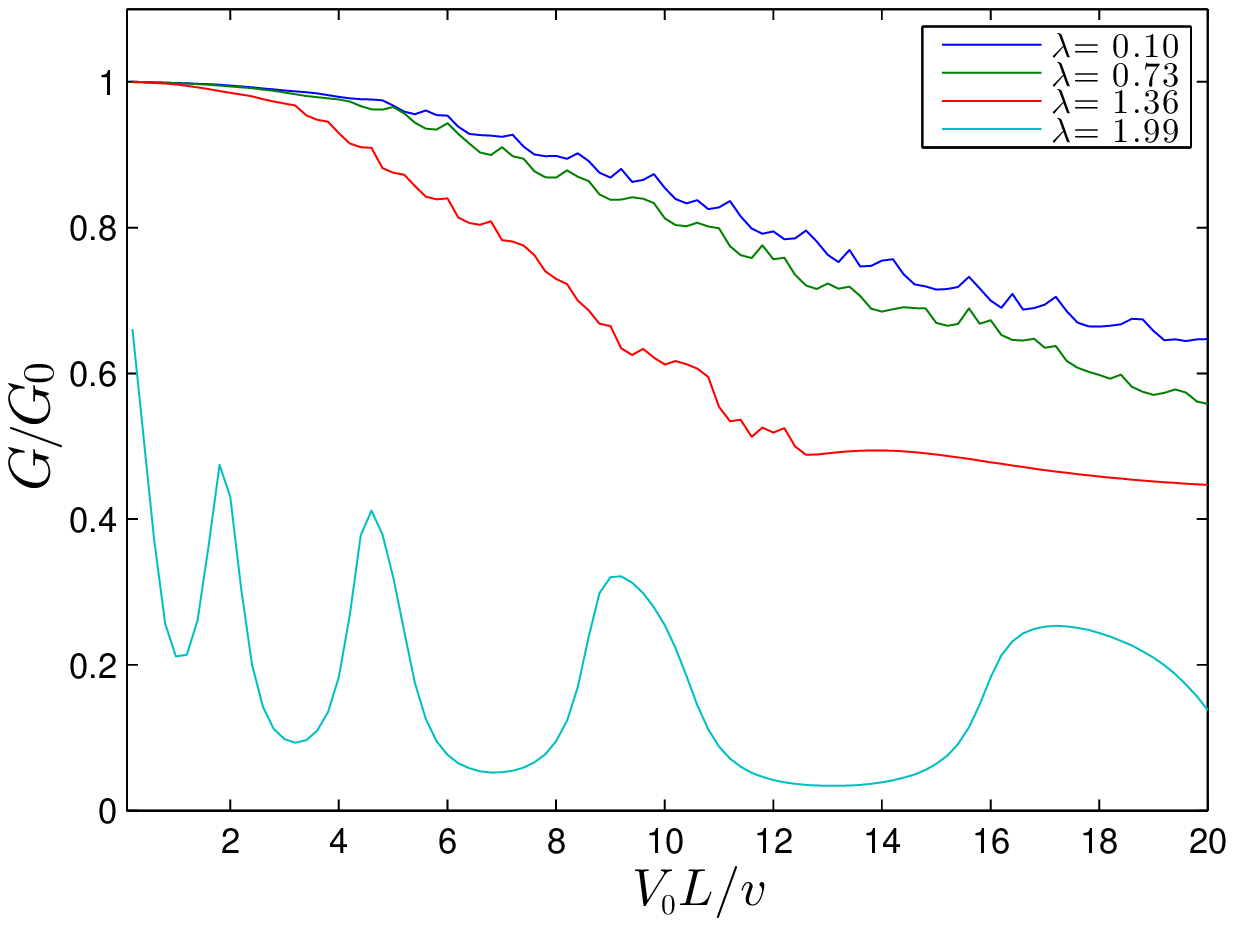}} 
\subfigure[]{\includegraphics[scale=0.4]{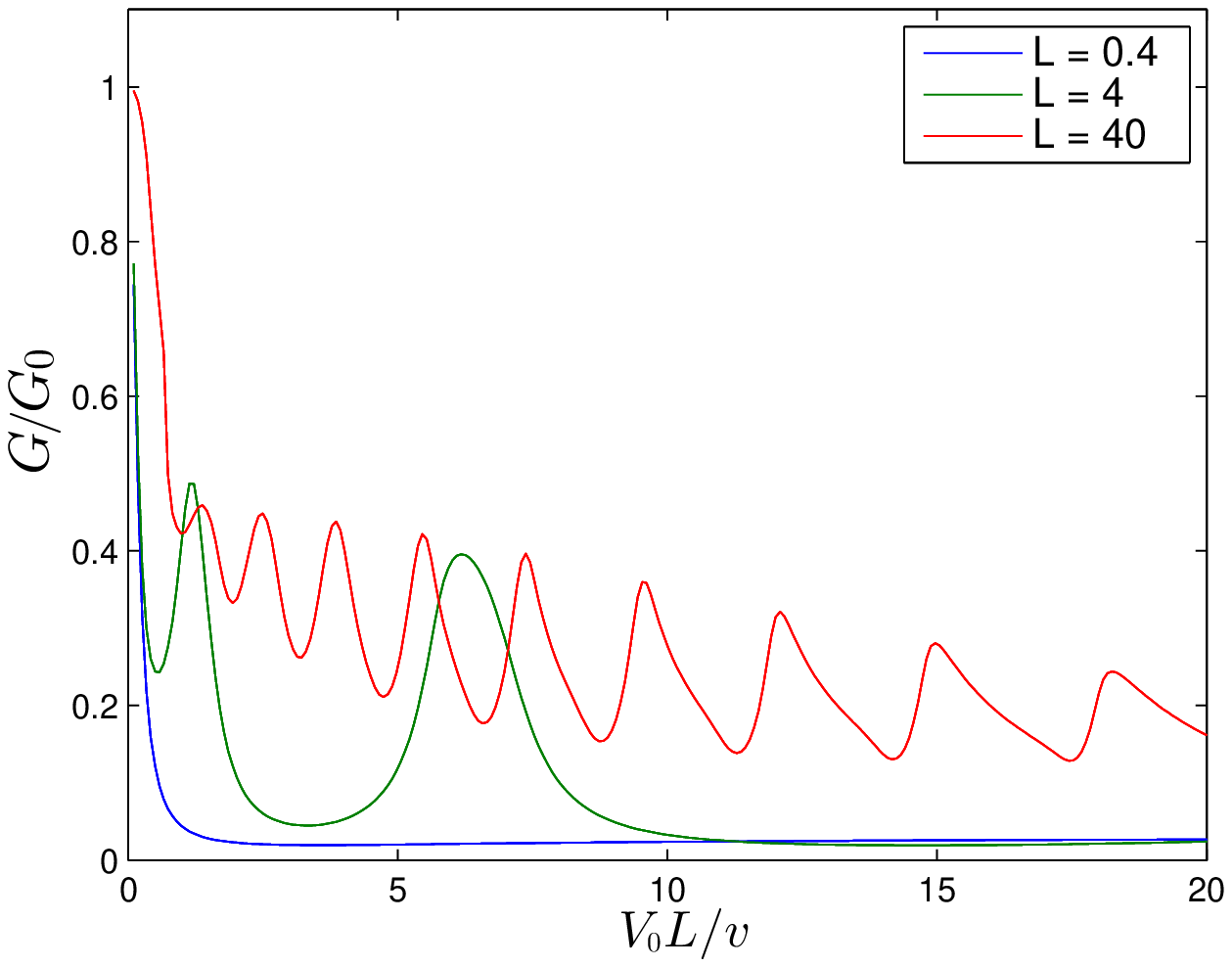}} 
\caption{Conductance versus $V_0 L/v$ for (a) different values of $\lm$ and
$L=1$, (b) different values of $\lm$ and $L=10$, and (c) different values of
$L$ and $\lm = 1.99$. We have taken $E=2$.} \label{fig12} 
\end{figure}
\end{center}
\end{widetext}

\subsection{Currents at top and bottom surfaces}
\label{sec6b}

We have again studied the transmitted currents and conductances
on the top and bottom surfaces separately.
We find that, just like the case of a potential barrier, the currents in
either of the surfaces can take negative values for certain values of $\ta$
when only one incident wave is present. When both incident waves are present,
we find that the current is always positive on both the surfaces.

\begin{figure}[H]
\centering
\hspace*{-.4cm} \subfigure[]{\includegraphics[scale=0.52]{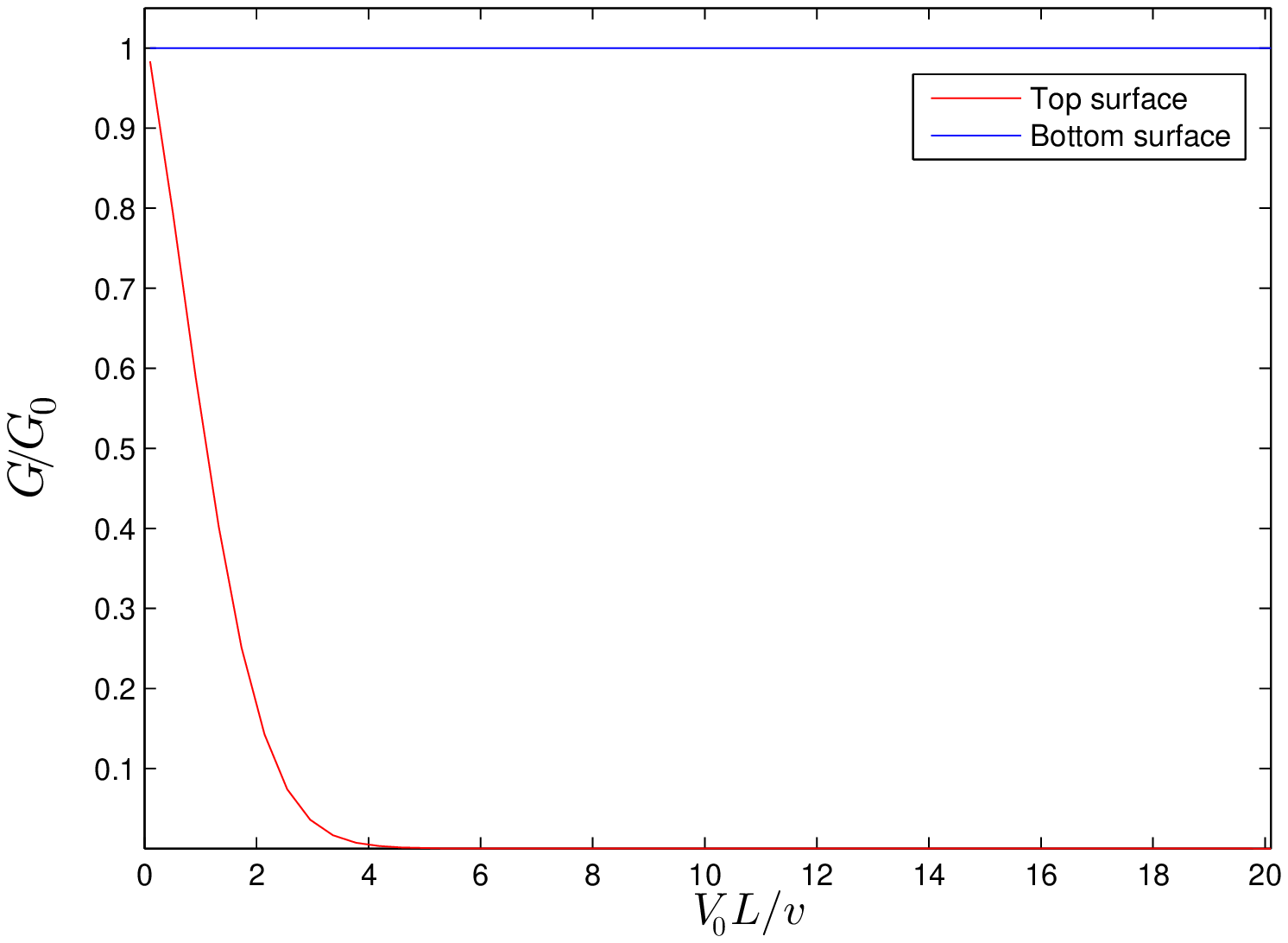}} 
\hspace*{-.4cm} \subfigure[]{\includegraphics[scale=0.62]{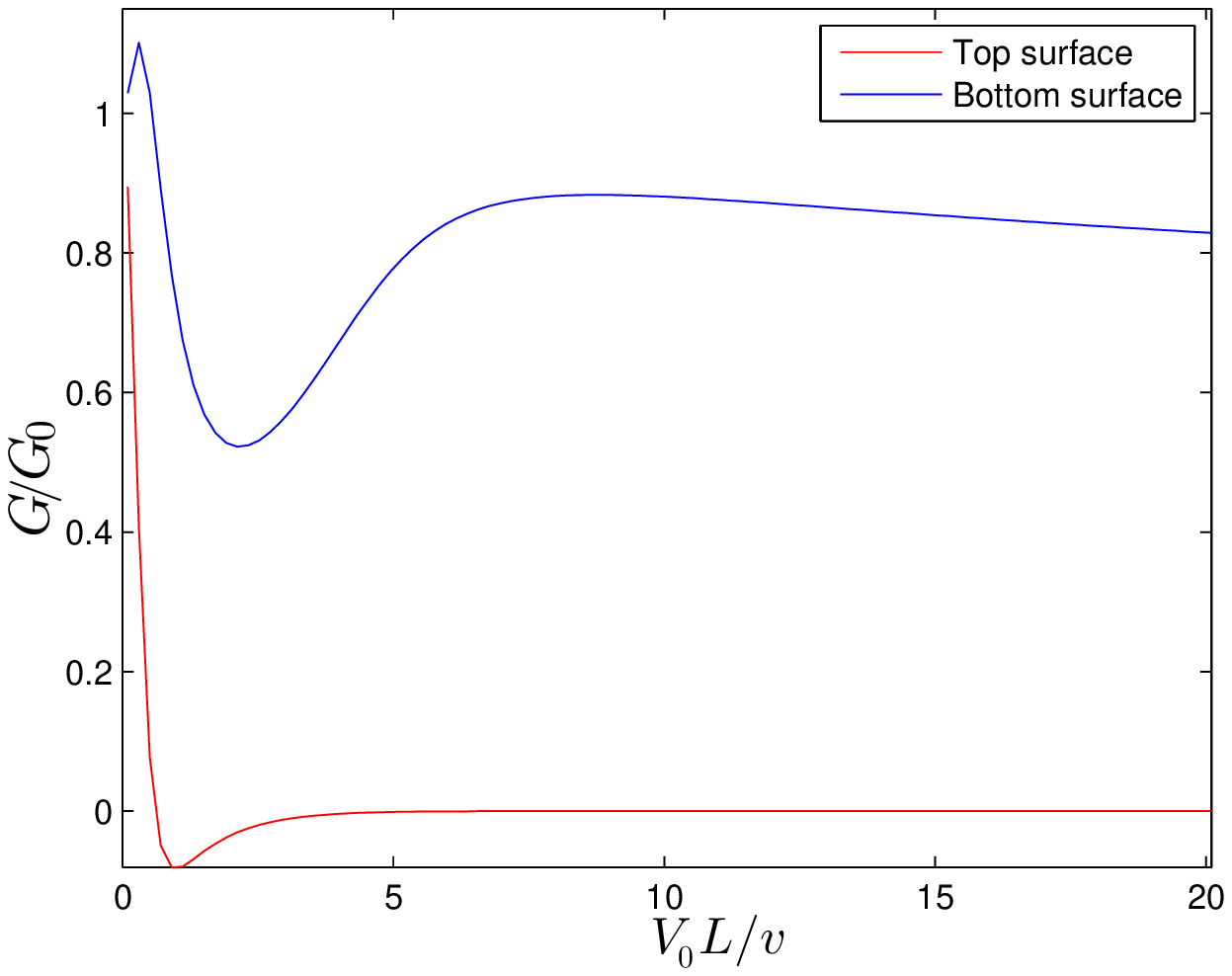}} 
\caption{Conductances at the top and bottom surfaces versus $V_0 L/v$ for
(a) a small value of $\lm= 0.01$ and (b) a value of $\lm= 1.9$ close to $E$.
We have taken $E=2$ and $L=1$.} \label{fig13} 
\end{figure}

The conductances at the top and bottom surfaces as a function of $V_0 L/v$ are
shown in Fig.~\ref{fig13} for two values of $\lm$.
Figure~\ref{fig13} (a) shows that when the
coupling $\lm$ is small, the bottom surface (which has no magnetic barrier)
conducts almost the same current for different values of the barrier
strength $V_0$, while the current at the top current decreases quickly as
$V_0$ increases. When the coupling $\lm$ has a value close to the energy $E$ 
(Fig.~\ref{fig13} (b)), the current at the top and bottom surfaces mix
producing a more complex behavior.
The current at the bottom surface decreases up to about $V_0 L/v = 2$ beyond
which it increases and reaches a constant. The current at the top surface
decreases up to about $V_0 L/v = 1$ where it is negative; beyond that value
it increases and eventually reaches a constant value of zero.
We note that this nonmonotonic variation with $V_0 L/v$ occurs only when the 
barrier width is substantial; in contrast, the behavior is monotonic for a 
$\de$-function magnetic barrier (Fig.~\ref{fig04}) or when the width is 
$0.4$ (Fig.~\ref{fig12} (c)).

\begin{figure}[H]
\centering
\hspace*{-.4cm} \includegraphics[scale=0.62]{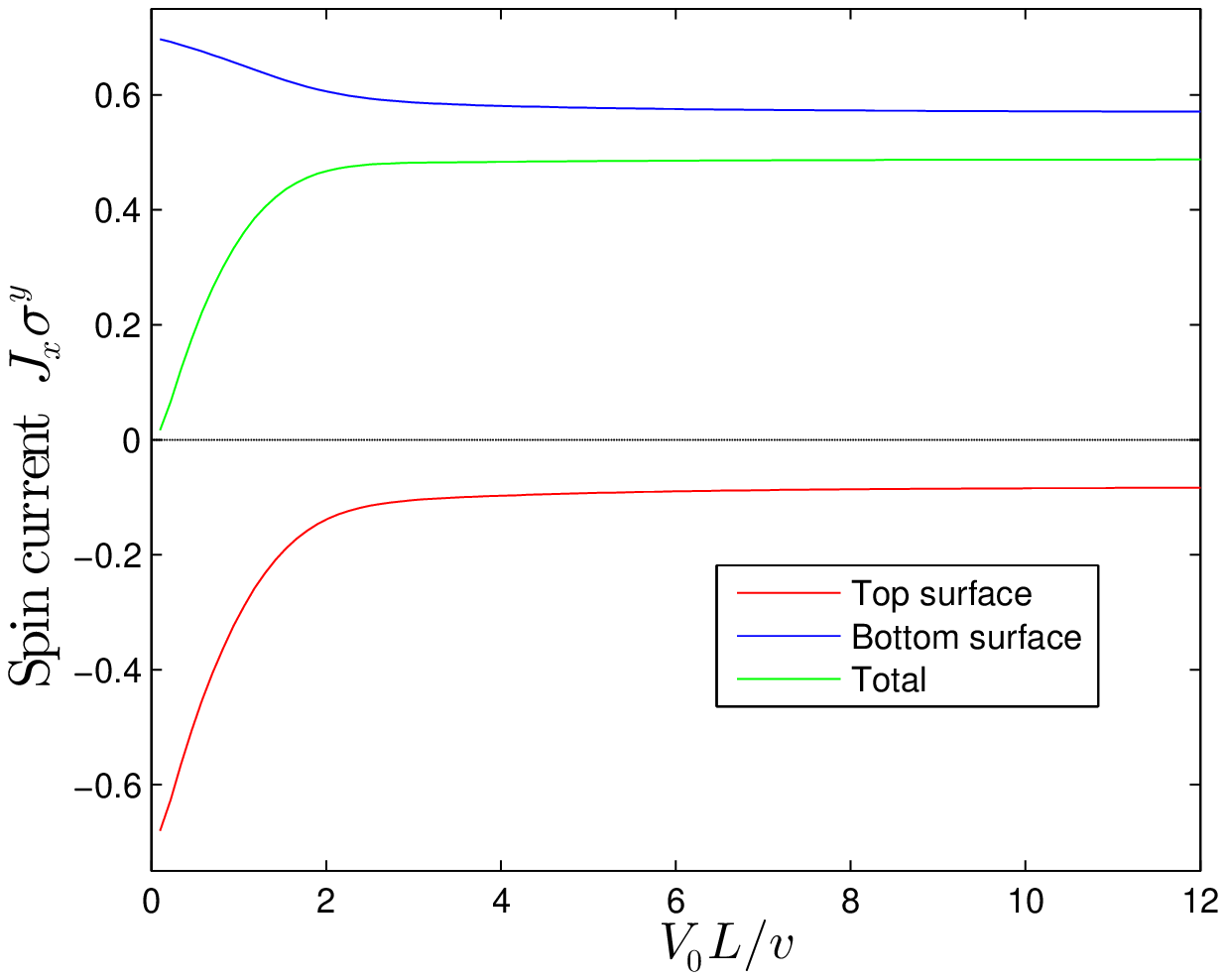} 
\caption{Total spin current (in units of $v$) as a function of $V_0L/v$ when 
both incident waves are present, $E=2$, $\lm =1$, and $L=1$.} \label{fig14} 
\end{figure}

Finally, we present plots of the transmitted spin current, similar to the
case of a potential barrier.
In Fig.~\ref{fig14}, we show the total spin current as a function of $V_0 L/v$
when both incident waves are present. We do not see any oscillations in the
spin current for the values of $\lm$ and $L$ chosen in this figure.
Indeed Fig.~\ref{fig14} looks very similar to Fig.~\ref{fig05} which 
showed the total spin current for a $\de$-function magnetic barrier.

\subsection{Magnetic barrier with other orientations of magnetization}
\label{sec6c}

We have so far studied the effects of a magnetic barrier in which the 
magnetization points along the $x$-direction. We will now discuss briefly 
what happens if the magnetization points along the $y$- or $z$-direction.
To obtain a qualitative understanding of these two cases, let us consider 
a $\de$-function magnetic barrier on the top surface similar to the
situation studied in Secs.~\ref{sec4c} and \ref{sec4d}.
If the magnetization points along the $y$-direction, we get a
Hamiltonian and a matching condition on the top surface given by
\beq H ~=~ v (-i\si^{x}\pa_{y} + i \si^{y}\pa_{x}) ~+~ c ~\de (x) ~\si^{y}, 
\eeq
and
\beq \psi_{x\to 0^{+}} ~=~ e^{i(c/v)} \psi_{x\to 0^{-}}. \eeq
This resembles the matching condition given in Eq.~\eqref{cond1} for a 
$\de$-function potential barrier in the sense that $\psi_{x\to 0^{+}}$ and
$\psi_{x\to 0^{-}}$ are related by a unitary transformation. We find 
numerically as well that the dependence of the conductance on
the various parameters is similar to the case of a $\de$-function potential
barrier. For instance, in both cases, the conductance oscillates with
increasing barrier strength $c$ as in Fig.~\ref{fig02}.

On the other hand, if the magnetization points along the $z$-direction, 
the Hamiltonian and matching condition on the top surface are given by
\beq H ~=~ v (-i\si^{x}\pa_{y} + i \si^{y}\pa_{x}) ~+~ c ~\de (x) ~\si^{z},
\eeq
and
\beq \psi_{x\to 0^{+}} ~=~ e^{- (c/v) \si^x} \psi_{x\to 0^{-}}. \eeq
This resembles the matching condition given in Eq.~\eqref{cond2} for a
$\de$-function magnetic barrier with magnetization pointing in that the 
$x$-direction in that the matrix connecting $\psi_{x\to 0^{+}}$ to 
$\psi_{x\to 0^{-}}$ is not unitary. Numerical calculations show that the 
dependence of the conductance on the various parameters is indeed similar to 
the case of a $\de$-function magnetic barrier with magnetization in the
$x$-direction. In both cases, the conductance becomes small and saturates
at a nonzero value with increasing $c$ as in Fig.~\ref{fig04}.

Thus the effects of a magnetic barrier with magnetization along the
$y$- and $z$-directions are, respectively, similar to a potential barrier and 
to a magnetic barrier with magnetization in the $x$-direction. \\

\section{Discussion}
\label{sec7}

In this work, we have studied a three-dimensional
topological insulator in which the states at the top and bottom
surfaces are coupled to each other, with the coupling being
characterized by an energy scale $\lm$. For each value of the energy
and surface momentum, there are two possible states which are linear
combinations of states at the top and bottom surfaces. We have
considered two types of barriers applied to the top surface, a
potential barrier and a magnetic barrier. We have studied the transmitted 
currents and conductances as functions of various parameters of the system: 
the angle of incidence $\ta$ of the incident waves, the coupling $\lm$, and 
the strength of the barrier $V_{0}$. We also studied the transmitted currents 
at the top and bottom surfaces separately which gives a clearer picture of the
contributions from the two surfaces. Further, we have studied the transmitted 
spin currents at the two surfaces separately. We note that the qualitative 
features of many of the results obtained for barriers with finite widths can 
be analytically understood using models with $\de$-function barriers. 

The main results obtained for potential barriers are as follows.
First, we have shown that the transmitted currents from the two
possible incident waves as a function of the angle of incidence
$\ta$ are symmetric about normal incidence ($\ta = \pi/2$). Moreover,
the conductance $G/G_0$ is, expectedly, an oscillatory function
of the barrier strength $V_0$. The difference of these
oscillations from their single surface counterpart is that their
period increases from $\pi$ to $2\pi$ (in dimensionless units) as we
increase the coupling $\lm$. The conductance at the peaks of these
oscillations reaches almost unity, independent of the value of $\lm$, for 
specific values of the barrier potential $V_0$ thus demonstrating near-perfect 
transmission resonances. Second, for a fixed value of $V_0$, the conductance 
as a function of the coupling $\lm$ decreases with increasing $\lm$. 
Third, looking at the currents at the top and bottom surfaces
separately, we find that when we send only one of the two possible
incident waves, the currents can take negative values for a small
range of values of $\ta$ close to glancing angles. This shows that
due to the coupling $\lm$ between the two surfaces, some current can
tunnel from the top surface to the bottom surface or vice versa. However, 
the sum of the currents when both incident waves are present is always
positive at both the surfaces. Fourth, the transmitted spin current
(with spin component along the $\hat y$ direction) is observed to be
always negative (positive) at the top (bottom) surface, but their
sum is always positive. This is due to the opposite forms of
spin-momentum locking on the two surfaces as mentioned after
Eq.~\eqref{hamb}; an electron with positive energy and moving in the
$+ \hat x$ direction on the top (bottom) surface has a spin pointing
in the $- \hat y$ ($+ \hat y$) direction, respectively. We note that this 
allows the usage of these junctions as splitters of currents into two separate 
spin currents with opposite polarizations. These spin currents can be picked 
up by attaching spin-polarized metallic leads to the two surfaces.

Next we summarize our main results for magnetic barriers. First, for
a barrier in which the magnetization points along the $\hat x$
direction, the transmitted current as a function of the angle of
incidence $\ta$ is not symmetric about normal incidence ($\ta =
\pi/2$), unlike the case of a potential barrier. This is because the
magnetic barrier breaks the symmetry $\ta \to \pi - \ta$. Moreover,
the normalized conductance $G/G_0$ does not oscillate but decreases
and reaches a constant value as the barrier strength $V_{0}$ increases, in
contrast to the case of a potential barrier. Even for very large
$V_{0}$, there is always a nonzero current due to the presence of
the bottom surface. Second, the conductance decreases as a function
of $\lm$ for a given value of $V_{0}$. As $E \to \lm$, the current
goes to zero. Third, the currents at the top and bottom surfaces
separately can again exhibit negative values near the glancing
angles, for the same reasons as mentioned above. Finally, the
transmitted spin currents have opposite signs on the top and
bottom surfaces due to the spin-momentum locking as discussed above.

In this work, we have not considered the effects of disorder. 
In the limit of strong nonmagnetic disorder, where the mean free path of 
the Dirac electrons becomes less than the width of the potential or
magnetic barrier, the effect of the disorder would have to be considered. 
This is, by itself, an interesting problem and could be a topic of future 
study. However, in this paper, we have concentrated on the other (ballistic) 
limit, where the mean free path of the Dirac electrons is much larger than 
the barrier width. In this weak disorder or ``clean" limit, as also pointed 
out in Ref.~\onlinecite{fogler} in the context of two-dimensional Dirac 
electrons in graphene, the transmission is not significantly affected. 
Further, such systems with weak disorder are experimentally feasible; thus 
this limit is expected to have experimental relevance.

The experimental verification of our results would involve transport
measurements in these systems. The best experimental set-up would 
involve four leads which separately connect to the top and bottom surfaces 
on the left and on the right of the barrier. One can then apply a common 
voltage to the two input leads on the left side from where the electrons 
are incident, and measure the currents individually at the two output leads 
on the right side where the electrons are transmitted. Apart from attaching 
the leads, one would also need to implement the
potential and magnetic barriers for these experiments. The potential
barriers can be implemented by putting gates across the top surface.
For magnetic barriers, one would need to deposit a layer of
magnetized material with magnetization along $\hat x$ on the top
surface; such a strip will induce a magnetization on the region below it
via the proximity effect and thus mimic the Hamiltonian of region $II$
\cite{mondal}. The first experiment that we suggest involves attaching 
spin-polarized leads with opposite spin polarizations, along $-\hat y$ and 
$+ \hat y$, at the top and bottom surfaces, respectively. This would allow one 
to pick up oppositely polarized spin currents as output for a generic charge 
current input in these junctions. We predict that a much smaller
output current will be picked up if the spin-polarizations of the leads on 
the two surfaces are reversed (i.e., $+ \hat y$ and $-\hat y$ at the top 
and bottom surfaces). Further, one can carry out a standard tunneling 
conductance measurement with these junctions in the presence of potential
barriers. The period of the oscillations of these tunneling
conductances as a function of the barrier strength (which could be
tuned using the gate voltage on the top surface) would depend on
$\lm/E$. Although it would be difficult to tune $\lm$, one can easily tune 
the incident electron energy $E$ and verify the change in the period of 
$G$ as a function of $\lm/E$ predicted in this work.

In conclusion, we have studied the transport across a junction of a thin 
topological insulator whose top and bottom surface are connected by a 
coupling of strength $\lm$ in the presence of either a potential or a
magnetic barrier atop its top surface. We have shown that such junctions show 
conductance oscillations as a function of the potential barrier strength whose 
period can be tuned by varying $E$. For a magnetic barrier, the conductance 
approaches a finite nonzero value with increasing barrier strength. 
We find that the spin currents on the top and bottom surfaces of such 
junctions always have opposite polarizations. Consequently, they can act as 
splitters of a charge current into two oppositely oriented spin currents. 
We have suggested experiments to test our theory.

\vspace{.8cm}
\centerline{\bf Acknowledgments}
\vspace{.5cm}

D.S. thanks DST, India for Project No. SR/S2/JCB-44/2010 for financial support.

\end{document}